\documentclass[review]{elsarticle}
\usepackage[left=0.6in, right=0.6in, top=1.2in, bottom=0.8in, includefoot, headheight=13.6pt]{geometry}
\usepackage{lineno,hyperref}
\modulolinenumbers[5]
\usepackage{amsmath}
\usepackage{amsfonts}
\usepackage{natbib}
\journal{Journal of \LaTeX\ Templates}

\begin{document}

\begin{frontmatter}

\title{Hamiltonian analysis and Faddeev-Jackiw formalism for two Dimensional Quadratic Gravity expressed as BF theory}

\author{Jaime Manuel Cabrera and Jorge Mauricio Paulin Fuentes}
\address{ Divisi\'on Acad\'emica de Ciencias B\'asicas, Universidad Ju\'arez Aut\'onoma de Tabasco,\\ 
 Km 1 Carretera
Cunduac\'an-Jalpa, Apartado Postal 24, 86690 Cunduac\'an, Tabasco, M\'exico}


            




\cortext[mycorrespondingauthor]{jaime.manuel@ujat.mx}


\begin{abstract}

We examine the model of Two-Dimensional Quadratic Gravity as a consequence of symmetry breaking within the framework of background field (BF) theory. This theory is essentially an extension of BF theory, introducing an additional polynomial term that operates on both the gauge and background fields. We analyze the theory using the Dirac and Faddeev-Jackiw procedures, determining the form of the gauge transformation, the full structure of the constraints, the counting of degrees of freedom, and the generalized Faddeev-Jackiw brackets. Additionally, we demonstrate the coincidence of the Faddeev-Jackiw and Dirac's brackets. Finally, we provide some remarks and discuss prospects.

\end{abstract}


\end{frontmatter}


\section{INTRODUCTION}

Throughout several decades, physicists have dedicated significant effort to generalize and extend General Relativity (GR) in various ways, with the goal of achieving a unified framework that encompasses both gravity and the other fundamental interactions within the framework of gauge theory, specifically Yang-Mills (YM) theory, as well as through the exploration of Topological Field Theories (TFT) \cite{Pe, Kras}. As previously mentioned, the search for a unified model of fundamental interactions has motivated the utilization of different frameworks. For example, one approach involves introducing higher dimensions in gravity models, where the metric remains the fundamental field and its components can be associated with the fundamental interactions \cite{Fair, Yuval}. On the other hand, the literature extensively highlights that elementary interactions can be comprehended by considering the connection associated with an internal symmetry group \cite{Wein}, where space-time is not considered a dynamic source. Numerous attempts have been made to construct Yang-Mills type theories of gravity. In this context, various formulations known as pure connection actions for gravity have been proposed \cite{Capo, Krasn, Ros, Mit, Cha}. The fundamental field in this perspective is the gauge field, which is associated with a specific symmetry group. Consequently, the metric no longer assumes the primary role in describing the gravitational field; instead, it becomes an object that can be derived from more fundamental elements. As a result, GR naturally emerges as a consequence of the proposed gauge theory. 
Another relevant approach to the unification of fundamental forces lies in the interplay between topological field theories and gravity. Within this framework, it has been discovered that GR can be formulated through a polynomial action principle that links it to the topological background field (BF) theory. This fact is quite surprising because, unlike BF theories, GR contains propagating degrees of freedom. In recent years, there has been considerable interest in the scientific community regarding the BF formulations of GR. These formulations have provided the groundwork for the covariant approach to quantum gravity called spinfoam models \cite{Ser}, which aim to quantize the gravitational field through a path integral method. Montesinos et al. \cite{Montesinos} have explored the different aspects of BF formulation in D-dimensions, covering a range of related models within the scope of GR. Despite the theoretical advances achieved through various alternative formulations to GR, a satisfactory quantum theory of gravity has not yet been established in the context of four-dimensional space-time. 
Therefore, we can resort to studying gravitational models in lower dimensions \cite{Brown} in order to gain a better understanding of some quantum properties of the gravitational field. In this regard, lower-dimensional BF models emerge as suitable environments for the analysis of quantum gravity in reduced dimensions (see \cite{Carlip} and references therein).  However, in lower-dimensional models, certain trivial features can be observed. For example, gravity in two dimensions cannot be explained by the framework of GR since, in this case, the theory becomes meaningless. However, Jackiw and Teitelboim \cite{Jackiw, Tei} suggested a model that enables the resolution of this problem by introducing a dilatonic field. On the other hand, the formulation of Jackiw-Teitelboim's approach as a BF gauge theory of $SO(2,1)$ has been developed in \cite{Fu}. There is another possibility to write down the action for 2-dimensional gravity in Polynomial BF form \cite{Pas}, using a like construction of the Freidel-Starodubsetv \cite{1, Smolin, Freidel}.
These models consist of two main components: a topological term involving assumed gauge symmetries, and a second quadratic term in the B fields that breaks the symmetry $SO(3)$. The action of GR originates from the introduction of symmetry breaking, accompanied by an additional term that is quadratic in curvature. The combination described gives the theory a structure that resembles modified gravity theories of the f(R) variety. These theories have the potential to provide an explanation for the recent acceleration of the universe, avoiding the need for the concept of dark energy \cite{Fe, No, Min}.  Additionally, there exists a range of compelling motives to explore secondary quadratic curvature Lagrangians within the domain of gravitational models. These motivations encompass several critical areas, including the analysis of quantum corrections \cite{Lan, Pais}, the resolution of divergences via renormalization techniques \cite{De,Hooft, Ste}, the examination of effective actions stemming from the framework of string theory \cite{Maro}, and the investigation of inflationary mechanisms that operate independently of the presence of scalar fields \cite{Staro}. 

Within the framework of the BF gauge approach to gravity, the theory is fundamentally defined by constraints and is characterized by the presence of gauge symmetries. In the context of gravity theories, this symmetry not only indicates that the physical content of the theory remains unchanged when the coordinates change, but it is also intricately related to the dynamics of the theory \cite{Bo}. When considering such systems, the standard method to study these theories is the Dirac-Bergmann method \cite{Dirac1, Dirac2, Dirac3, Ber, Ber1, Ber2}. In this approach, the Hamiltonian is a linear combination of constraints, and the dynamic behavior of the system is characterized by these constraints. This formalism categorizes constraints into primary, secondary, and tertiary classes, and distinguishes between first-class and second-class constraints. The presence of first-class constraints results in the emergence of gauge symmetries, whereas the existence of second-class constraints leads to a reduction in the number of phase space variables.

In this regard, achieving a precise characterization of the dynamic structure in Polynomial BF theories requires a thorough evaluation of the relevant constraints. Therefore, conducting a comprehensive analysis and careful consideration is necessary to identify these constraints. Until now, no comprehensive study of the canonical description and constraint analysis of the Polynomial BF theory proposed by Paszko-da Rocha \cite{Pas} in the full phase space has been conducted. However, an initial approach has been made in the reduced phase space and presented in \cite{Val}. It is worth noting that understanding the nature of constraints in the full phase space and their implications for the canonical description of the Polynomial  BF model is a matter of great importance in both classical and quantum gravity. The implementation of the Dirac-Bergmann formalism requires careful consideration, as obtaining and classifying the constraints within the formalism is not an easy task.
Within this context, an additional and powerful approach that has been developed for investigating canonical singular systems was introduced by Faddeev and Jackiw \cite{Fa}. In this formalism, all constraints are treated equally, without the need to impose any additional classification.  A central requirement of this approach is to have a first-order Lagrangian that describes the system. The constraints are derived from the zero modes of the symplectic two-form matrix and are gradually introduced into the Lagrangian until all the constraints in the theory are identified \cite{Bar, Bar1}. This approach has been extended through modifications to include the Dirac-Bergmann algorithm, as detailed in \cite{Long, Liao}.  It is noteworthy that the symplectic matrix holds crucial information within the theory. The inverse of this matrix yields the FJ generalized brackets, aligning with the Dirac brackets \cite{Liao, Gar}. 

The Faddeev-Jackiw formalism has been successfully implemented in various physical systems, including exotic actions for both Abelian and non-Abelian gravity \cite{Jaime}, four-dimensional BF theory \cite{Esca}, the Bonzom–Livine model for gravity \cite{Jaime2}, the Christ-Lee model \cite{Gupta}, a particle constrained to move in a torus knot \cite{An}, and the Jackiw-Teitelboim model as BF theory in two dimensions \cite{Jaime1}.

The primary goal of this article is to conduct a thorough analysis of the Polynomial BF action in 1+1 dimensions, as proposed by Paszko-da Rocha \cite{Pas}. We aim to achieve this by applying two classical methods commonly used in dealing with constrained systems. The initial approach we will employ is inspired by the contributions of Dirac and Bergmann. In this methodology, we treat all the variables within the theory as dynamic variables. This approach offers the distinct advantage of fully utilizing the phase space, rather than restricting our attention solely to variables explicitly linked with generalized velocities in the action. Furthermore, our analysis is conducted without fixing the gauge, and we exclusively eliminate second-class constraints by utilizing Dirac brackets. Consequently, we preserve the integrity of the first-class constraints. These unique characteristics differentiate our approach from the analysis presented in \cite{Val}, where this specific aspect was not taken into account. In the second description, the modified FJ approach simplifies the need to classify constraints into first and second classes, treating all constraints uniformly \cite{Long, Liao}. Within the framework of the modified Faddeev–Jackiw formalism (FJ), we explore the structure of constraints and the derivation of generalized FJ brackets. As a final step, we aim to obtain the gauge transformations.





The structure of this paper is organized as follows. In Section II, a comprehensive analysis is conducted on the model proposed in \cite{Pas}. This analysis involves an in-depth exploration of the relationship between the two-dimensional Polynomial BF action and quadratic gravity, employing a methodology closely aligned with that presented in \cite{Val}. Section III is dedicated to carry out a comprehensive Hamiltonian analysis, we will demonstrate that a thorough examination of the dynamical structure reveals the entire set of physical constraints, encompassing primary and secondary. 
In Section IV, our focus turns to identifying the first and second-class constraints within the Polynomial BF theory and determining the corresponding physical degrees of freedom in our model. Besides, we derive the extended action and Hamiltonian. Following this, we proceed to establish the fundamental quantization brackets. Finally, we delve into the derivation of gauge transformations and subsequently identify the generators associated with these transformations. Section V is devoted to a meticulous presentation of the Faddeev–Jackiw symplectic analysis of the Polynomial BF action. This analysis serves as the foundation for the determination of all physical constraints through the solution of zero-modes within the singular symplectic matrix. Additionally, we establish the core quantization brackets, articulate the functional measure in the path integral, and recognize the physical degrees of freedom linked to our model through the introduction of a gauge-fixing procedure.

\section{2D quadratic gravity expressed as a Polynomial BF action}


As mentioned in \cite{Pas}, the action of quadratic gravity can be represented as a 2D polynomial BF action. In this formulation, the process begins by constructing a BF action that remains invariant under $SO(3)$. Subsequently, a quadratic interaction term in A and B is introduced.
As a result of introducing this interaction term, the initial invariance is broken, leading to a reduction to $SO(2)$. Rather than introducing the interaction term directly through Lagrange multipliers, it was suggested to follow an approach akin to that proposed by MacDowell-Mansouri \cite{1}.


 The theory is formulated by means of 0-form field $B=\frac{1}{2}B^{IJ}T_{IJ}$, and the curvature\footnote{Where $F\equiv dA + A\wedge A=\frac{1}{4}(\partial_{\alpha}A^{IJ}_{\beta}-\partial_{\beta}A^{IJ}_{\alpha}+A^{IK}_{\alpha}A{_{\beta K}}^{J}-A^{IK}_{\beta}A{_{\alpha K}}^{J})T_{IJ}dx^{\alpha}\wedge dx^{\beta}$.} $F=\frac{1}{4}F^{IJ}_{\alpha\beta}T_{IJ}dx^{\alpha}\wedge dx^{\beta}$ written in terms of a 1-form field $A=A^{IJ}_{\mu}dx^{\mu}T_{IJ}$, where $T_{IJ}$ (with $I=0,1,2$) are the generators of the $SO(3)$ group and metric $\eta_{IJ}=diag(+,+,+)$.

The action of the BF model with interaction term
is given by \cite{Pas}



\begin{eqnarray}
S=\int_{\mathcal{M}} tr [-\frac{1}{2}B\wedge F + k^{2}(PB)(PB)(QA)\wedge A],
\label{1}
\end{eqnarray}

where $k$ is a constant parameter.
To develop the polynomial BF action (\ref{1}), the pivotal insight in reference \cite{Pas} was to treat P and Q as projection operators defined by\footnote{$\epsilon^{IJK}$ is the Levi-Civita symbol and, by convention we have $\epsilon^{012}=1$.}

\begin{eqnarray}
P^{IJ,KL}&\equiv&\frac{1}{2}\epsilon^{IJ2}\epsilon^{KL2},
\label{2} \\
Q_{IJ,KL}&\equiv&\frac{1}{2}\epsilon{_{IJ}}^{M}\epsilon{_{KL}}^{N}\epsilon_{MN2}.
\label{3}
\end{eqnarray}



Let P be a projection operator into the subalgebra of Lorentz generators $T_{ab}$ (with a = 0,1) and translations $T_{a2} = P_{a}$. It is evident that the operator (\ref{2}) does not project any translations; in other words, it does not affect the translations but does leave the Lorentz generators invariant

\begin{eqnarray}
(PT)_{ab}=T_{ab}, \quad (PT)_{a2}=0,
\label{4}
\end{eqnarray}
and the operator Q described in (\ref{3}) does not project any Lorentz generators; instead, it exchanges the translations

\begin{eqnarray}
(Q T)_{ab}=\epsilon{_{a}}^{b}T_{b2}, \quad (Q T)_{a2}=0.
\label{5}
\end{eqnarray}



Instead of employing differential forms, we work with the components of the B and F fields, resulting in \cite{Pas, Val}

\begin{equation}
S=\int d^{2}x\epsilon^{\mu\nu}\bigg[\frac{1}{4}B_{KL}F^{KL}_{\mu\nu}+\frac{k^{2}}{2}(B_{ab}B^{ab})\epsilon_{mn}A^{m}_{\nu}A^{n}_{\nu}\bigg].
\label{5a}
\end{equation}


where $\epsilon^{\mu\nu}$ represents the Levi-Civita Symbol for the spacetime. 



It has been demonstrated in \cite{Pas} that the action given in (\ref{5a}) is equivalent to a quadratic gravity model. To see this one ﬁrst decomposes the connection $A^{IJ}_{\mu}$ 
into zweiben $lA^{a2}_{\mu}=e^{a}_{\mu}$  and Lorentz connection  $A^{ab}_{\mu}=\omega^{ab}_{\mu}=\epsilon^{ab}\omega_{\mu}$ as follows:



\begin{eqnarray}
A_{\mu}&=&\frac{1}{2}A^{IJ}_{\mu}T_{IJ}=\frac{1}{2}A^{ab}_{\mu} T_{ab}+A^{a2}_{\mu}T_{a2}, \nonumber \\
       &=& \frac{1}{2}\omega^{ab}_{\mu}T_{ab}+ \frac{1}{l}e^{a}_{\mu}P_{a}.
\label{6}
\end{eqnarray}



The parameter $l$ is introduced for dimensional considerations, given that the zweibein is dimensionless. As we delve further, we will discover that this parameter is connected to the cosmological constant. Additionally, the components of the connection curvature $A^{IJ}$ are linked to the curvature of the Lorentz connection $\omega$


\begin{eqnarray}
F^{ab}_{\mu\nu}&=&
\epsilon^{ab}(\partial_{\mu}\omega_{\nu}-\partial_{\nu}\omega_{\mu})-\frac{1}{l^{2}}(e^{a}_{\mu}e^{b}_{\nu}-e^{a}_{\nu}e^{b}_{\mu})=R^{ab}_{\mu\nu}-\frac{1}{l^{2}}(e^{a}_{\mu}e^{b}_{\nu}-e^{a}_{\nu}e^{b}_{\mu}),
\label{7}
\end{eqnarray}

and the torsion

\begin{eqnarray}
F^{a2}_{\mu\nu}&=& \frac{1}{l}(\partial_{\mu}e^{a}_{\nu}-\partial_{\nu}e^{a}_{\mu}+\epsilon{^{a}}_{b}\omega_{\mu}e^{b}_{\nu}-\epsilon{^{a}}_{b}\omega_{\nu}e^{b}_{\mu})=\frac{1}{l}T^{a}_{\mu\nu}.
\label{7a}
\end{eqnarray}



By utilizing equations (\ref{6}) and (\ref{7}), as well as (\ref{7a}), we can express the action of the theory in the polynomial BF action (\ref{5a}) in the following way \footnote{The following notations are used $e= det e^{a}_{\mu}$ is the determinant of the zweiben, and $B_{a}=B_{a2}$,
$B_{ab}=\epsilon_{ab}B$.}


\begin{eqnarray}
S[e^{a}_{\mu},\omega_{\mu},B,B_{a}]=\int d^{2}x \bigg[\frac{1}{2}B\bigg(\frac{1}{2}\epsilon_{ab}\epsilon^{\mu\nu}R^{ab}_{\mu\nu}-\frac{2}{l^{2}}e\bigg)+\frac{1}{2l}\epsilon^{\mu\nu}B_{a}T^{a}_{\mu\nu}+\frac{2k^{2} }{l^{2}}B^{2}e\bigg].
\label{9}
\end{eqnarray}





By varying the action (\ref{9}) with respect to $(e^{a}_{\mu}, \omega_{\mu}, B, B_{a})$, the following equations of motion are derived

\begin{eqnarray}
\frac{\delta S}{\delta e^{a}_{\mu}}&:& \frac{1}{l}\epsilon^{\mu\nu}(D_{\nu}B_{a}-\epsilon_{ab}\frac{B}{l}e^{b}_{\nu}(1-2k^{2}B))=0,
\label{9a} \\
\frac{\delta S}{\delta\omega_{\mu}}&:& \epsilon^{\mu\nu}(\partial_{\nu}B+\frac{1}{l}\epsilon_{ab}B^{a}e^{b}_{\nu})=0, \label{9b}\\
\frac{\delta S}{\delta B}&:& \frac{1}{4}\epsilon_{ab}\epsilon^{\mu\nu}R^{ab}_{\mu\nu}-\frac{e}{l^{2}}(1-4k^{2}B)=0, \label{9c}\\
\frac{\delta S}{\delta B_{a}}&:& \epsilon^{\mu\nu}T^{a}_{\mu\nu}=0.
\label{10}
\end{eqnarray}




Furthermore, as a special case of the Lagrangian (\ref{9}), the Jackiw-Teitelboim model emerges when we set $k=0$ in equation (\ref{9c}). This choice results in the elimination of an interaction term. Notably, in this scenario, the B field retains its arbitrariness, allowing for the establishment of an identification with the dilatonic scalar field. Additionally, by selecting the cosmological constant $\Lambda=1/l^{2}$ and defining $R = g^{\mu\nu}R_{\mu\nu}$ as the curvature scalar, the action (\ref{9}) seamlessly transforms into the Jackiw-Teitelboim action \cite{Jackiw, Tei}

\begin{eqnarray}
S_{JT}=\frac{1}{2}\int \sqrt{g}B(R-2\Lambda),
\label{12}
\end{eqnarray}




On the other hand, in the region where $k\neq 0$, we can use equation (\ref{9c}) to solve for the field B, and we can establish a relationship between the B field, curvature, and the parameter $l^{2}$  \cite{Pas, Val}

\begin{eqnarray}
B=-\frac{l^{2}}{8k^{2}} (R-\frac{2}{l^{2}}).
\end{eqnarray}

and then substitute it into action (\ref{9}) in order to obtain\footnote{where $G=k^{2}/8\pi$ and $2\Lambda = 1/l^{2}$}


\begin{eqnarray}
S=\frac{1}{16\pi G }\int d^{2}x \sqrt{g}(R-2\Lambda)-\frac{1}{128\pi G \Lambda}\int  d^{2}x \sqrt{g}R^{2}.
\label{11}
\end{eqnarray}

Hence, the two-dimensional Polynomial BF action (\ref{1}) can be regarded as equivalent to a quadratic gravitational action, at least in the classical sense.

\section{Hamiltonian Constraint Analysis}

Hence, in order to develop the Hamiltonian analysis, we assume for the space-time a topological structure of the form $M=\Re\times\Sigma$, where the real line $\Re$ represents an evolution parameter, and $\Sigma$ is a 1-dimensional manifold arbitrary but fixed topology, representing the "space". By performing the 1+1 decomposition, we can write the action (\ref{9}) as\footnote{Covariant derivatives $D_{1}B_{a}=\partial_{1}B_{a}+\epsilon{_{a}}^{b}\omega_{1}B_{b}$.}

\begin{eqnarray}
S[e^{a}_{\mu},\omega_{\mu},B,B_{a}]= \int d^{2}x \bigg(B\dot{\omega}_{1}+\frac{1}{l}B_{a}\dot{e}^{a}_{1}+\omega_{0}(\partial_{1}B+\frac{1}{l}\epsilon{^{a}}_{b}B_{a}e^{b}_{1})+\frac{e^{a}_{0}}{l}(D_{1}B_{a}-\frac{1}{l}\epsilon_{ab}Be^{b}_{1}(1-2k^{2}B))\bigg),
\nonumber\\
\label{13}
\end{eqnarray}




The definition of the momenta $\Pi^{Q_{L}}=(\Pi,\Pi^{a},\Pi^{\mu},\Pi^{\mu}_{a})$ canonically conjugate to the configuration variables $Q_{L}=( B, B_{a}, \omega_{\mu}, e^{a}_{\mu})$ are determined by
\begin{equation}
\Pi= \frac{\delta {\mathcal{L}} }{ \delta \dot{B} },  \qquad \Pi^{a}= \frac{\delta {\mathcal{L}} }{ \delta \dot{B}_{a} }, \qquad  \Pi^{\mu}= \frac{\delta {\mathcal{L}} }{ \delta \dot{\omega}_{\mu} }, \qquad  \Pi^{\mu}_{a}= \frac{\delta {\mathcal{L}} }{ \delta \dot{e}^{a}_{\mu} }.
\label{16}
\end{equation}
The matrix elements of the Hessian is given by
\begin{equation}
 H_{LM}=\frac{\partial^2{\mathcal{L}} }{\partial \dot{Q}_{L} \partial \dot{Q}_{M}}=0.
\label{17}
\end{equation}

Notice that the rank of the Hessian is zero, thus, we expect $9$ primary constraints. From the definition of the momenta $(\ref{16})$ we identify to the following primary constraints


\begin{eqnarray}
\phi:= \Pi \approx 0,\quad
\phi^{a}:= \Pi^{a} \approx 0 , \quad
 \phi^{0}:= \Pi^{0} \approx 0, \quad
 \phi^{1}:= \Pi^{1}- B\approx 0, \quad \phi^{0}_{a}:= \Pi^{0}_{a}\approx 0, \quad \phi^{1}_{a}:= \Pi^{1}_{a}- \frac{1}{l}B_{a}\approx 0.
\label{18}
\end{eqnarray}

Due to the constraints (\ref{18}) the motion is restricted to a subspace $\Gamma_{1}$ of the full phase-sapce $\Gamma$. The Phase-space $\Gamma$ of this system includes the dynamical fields $(Q_{L},\Pi^{Q_{L}})$.

The two remaining terms in the action (\ref{13}) give the canonical Hamiltonian and  $H_{c}$ is only defined on $\Gamma_{1}$ 
\begin{equation}
H_c= -\int dx \bigg(\omega_{0}(\partial_{1}B+\frac{1}{l}\epsilon{^{a}}_{b}e^{b}_{1}B_{a})+\frac{e^{a}_{0}}{l}(D_{1}B_{a}-\frac{1}{l}\epsilon_{ab}e^{b}_{1}B(1-2k^{2}B))\bigg),
\label{19}
\end{equation}



this expression, independent of velocities, solely encompasses spatial derivatives of coordinates and momenta. Additionally, $H_{c}$ can be considered as the restriction to the hypersurface $\Gamma_{1}$ of a function $H_{P}$, which we shall call primary Hamiltonian defined all over phase-space. Upon incorporating the primary constraints (\ref{18}), the primary Hamiltonian takes on the following form

\begin{equation}
H_P= H_c + \int dx \left[  \check{\lambda}\phi + \check{\lambda}_{a}\phi^{a} + \check{\lambda}_{\mu}\phi^{\mu}+\check{\lambda}^{a}_{\mu}\phi^{\mu}_{a} \right],
\label{20}
\end{equation}
where $ (\check{\lambda}, \check{\lambda}_{a}, \check{\lambda}_{\mu},\check{\lambda}^{a}_{\mu})$ are the corresponding  Lagrange multipliers associated of these constraints $(\phi,\phi^{a},\phi^{\mu},\phi^{\mu}_{a})$. The fundamental Poisson brackets of the theory  are determined by the commutation relations
\begin{eqnarray}
\{ B(x),\Pi(y) \}  &=& \delta(x-y), \nonumber \\
\{ B_{a}(x), \Pi^{b}(y) \}  &=& \delta^{b}_{a} \delta(x-y), \nonumber \\
\{ e^{a}_{\mu}(x), \Pi^{\nu}_{b}(y) \}  &=& \delta^{a}_{b} \delta^{\nu}_{\mu} \delta(x-y), \nonumber \\
\{ \omega_{\mu}(x),\Pi^{\nu}(y) \}  &=& \delta^{\nu}_{\mu} \delta(x-y).
\label{21}
\end{eqnarray}

The next step is to observe if there are more constraints, so that, we calculate the following 9$\times$9 matrix whose entries are the Poisson brackets among the constraints (\ref{18}). The non-vanishing Poisson
brackets are given by
\begin{eqnarray}
\{ \phi (x), \phi^{1} (y) \} &=& \delta(x-y), \nonumber \\
\{ \phi^{a} (x), \phi^{1}_{b} (y) \} &=& \frac{1}{l}\delta^{a}_{b}\delta(x-y),
\label{22}
\end{eqnarray}

and expressed in matrix form, namely, $W=\{\widetilde{\Phi}^{A}(x),\widetilde{\Phi}^{B}(y)\}$


\begin{eqnarray}
\left[ 
\begin{array}{l|cccccc}
       & \phi(y)    & \phi^{b}(y)    & \phi^{0}(y) & \phi^{1}(y) & \phi^{0}_{b}(y) & \phi^{1}_{b}(y)    \\ \hline
\phi(x)   & 0& 0 & 0 &\{ \phi(x), \phi^{1}(y) \} & 0 & 0  \\
\phi^{a}(x)    & 0&0&0 & 0& 0  & \{ \phi^{a}(x), \phi^{1}_{b}(y) \}\\
\phi^{0}(x)    & 0&0&0 & 0& 0  & 0\\
\phi^{1}(x)    &\{ \phi^{1}(x), \phi(y) \} &0&0 & 0& 0  & 0\\
\phi^{0}_{a}(x)    &0 &0&0 & 0& 0  & 0\\
\phi^{1}_{a}(x)   & 0 &\{ \phi^{1}_{a} (x), \phi^{b} (y) \}&0&0&0&0\\
\end{array}
\right]
,
\label{22a}
\end{eqnarray}

where, $ \widetilde{\Phi}^{A}=(\phi,\phi^{a},\phi^{\mu},\phi^{\mu}_{a})$. We can see that this matrix has rank=6 and 3 null-vectors. By using these 3 null-vectors and the evolution of $\phi^{0}_{a} $ and $\phi^{0}$ produces the following  $3$ secondary constraints
\begin{eqnarray}
\dot{\phi}^{0}&=& \{\phi^{0} (x), {H}_{P} \} \approx 0 \quad \Rightarrow \quad \Psi:= \partial_{1}B+\frac{1}{l}\epsilon{^{a}}_{b}e^{b}_{1}B_{a}\approx 0, \nonumber \\
\dot{\phi}^{0}_{a}&=& \{\phi^{0}_{a} (x), {H}_{P} \} \approx 0 \quad \Rightarrow \quad \Psi_{a}:= D_{1}B_{a}-\frac{1}{l}\epsilon_{ab}e^{b}_{1}B(1-2Bk^{2})\approx 0,
\label{23}
\end{eqnarray}


and the rank enables us to determine the specific values for the Lagrangian multipliers

\begin{eqnarray}
\dot{\phi}&=&\{\phi(x),H_{P}\}\approx0 \quad \Rightarrow \check{\lambda}_{1}-\partial_{1}\omega_{0}-\frac{1}{l^{2}}\epsilon_{ab}e^{b}_{1}e^{a}_{0}(1-4k^{2}B)\approx0,\nonumber \\
\dot{\phi}^{a}&=&\{\phi^{a}(x),H_{P}\}\approx0 \quad \Rightarrow \check{\lambda}^{a}_{1}-\partial_{1}e^{a}_{0}-\epsilon{^{a}}_{b}\omega_{1}e^{b}_{0}+\epsilon{^{a}}_{b}e^{b}_{1}\omega_{0}\approx0, \nonumber \\
\dot{\phi}^{1}&=&\{\phi^{1}(x),H_{P}\}\approx0 \quad \Rightarrow \check{\lambda}-\frac{1}{l}\varepsilon{_{a}}^{b}B_{b}e^{a}_{0} \approx 0, \nonumber \\
\dot{\phi}^{1}_{a}&=&\{\phi^{1}_{a}(x),H_{P}\}\approx0 \quad \Rightarrow \check{\lambda}_{a}+\epsilon{_{a}}^{b}B_{b}\omega_{0}-\frac{1}{l}\epsilon_{ab}e^{b}_{0}B(1-2k^{2}B)\approx0.
\label{24}
\end{eqnarray}



Consistency demands the preservation of the secondary constraints over time. In this theory, there are no third-order constraints.

\section{First- and second-class constraints}

\subsection{Separation of constraints}

To categorize the full set of constraints, we must distinguish between first and second-class constraints by identifying them among the primary and secondary constraints. To achieve this goal, we must calculate the rank and null-vectors of a 12x12 matrix. The matrix's elements will consist of the Poisson brackets between primary and secondary constraints. This process involves the following steps

\begin{eqnarray}
\{ \phi (x), \phi^{1} (y) \}&=& \delta(x-y), \nonumber \\
\{ \phi (x), \Psi (y) \}&=& -\partial_{1}\delta(x-y), \nonumber \\
\{ \phi (x), \Psi_{b} (y) \}&=& \frac{1}{l}\epsilon_{bc}e^{c}_{1}(1-4Bk^{2})\delta(x-y), \nonumber \\
\{ \phi^{a} (x), \phi^{1}_{b} (y) \}&=& \frac{1}{l}\delta^{a}_{b}\delta(x-y), \nonumber\\
\{ \phi^{a} (x), \Psi (y) \}&=& -\frac{1}{l}\epsilon{^{a}}_{b}e^{b}_{1}\delta(x-y), \nonumber\\
\{ \phi^{a} (x), \Psi_{b} (y) \}&=& -\delta^{a}_{b}\partial_{1}\delta(x-y)+\epsilon{^{a}}_{b}\omega_{1}\delta(x-y), \nonumber \\
\{ \phi^{1} (x), \Psi_{b} \}&=& -\epsilon{_{b}}^{c}B_{c},  \nonumber \\
\{\phi^{1}_{a} (x),\Psi (y)\}&=&  -\frac{1}{l}\epsilon{^{b}}_{a}B_{b}\delta(x-y), \nonumber\\
\{\phi^{1}_{a} (x),\Psi_{b} (y)\}&=& -\frac{1}{l}B\epsilon_{ab}(1-2Bk^{2})\delta(x-y).
\label{25}
\end{eqnarray}

Furthermore, it can be represented in matrix form as follows


\begin{eqnarray*}
\left[ 
\begin{array}{l|cccccccc}
       & \phi(y)    & \phi^{b}(y)    & \phi^{0}(y) & \phi^{1}(y) & \phi^{0}_{b}(y) & \phi^{1}_{b}(y) & \Psi(y) & \Psi_{b}(y)   \\ \hline
\phi(x)   & 0& 0 & 0 &\{ \phi(x), \phi^{1}(y) \} & 0 & 0 & \{ \phi(x), \Psi(y) \}& \{ \phi(x), \Psi_{b}(y) \}\\
\phi^{a}(x)    & 0&0&0 & 0& 0  & \{ \phi^{a}(x), \phi^{1}_{b}(y) \}&\{ \phi^{a}(x), \Psi(y) \}& \{ \phi^{a}(x), \Psi_{b}(y) \}\\
\phi^{0}(x)    & 0&0&0 & 0& 0  & 0&0&0\\
\phi^{1}(x)    &\{ \phi^{1}(x), \phi(y) \} &0&0 & 0& 0  & 0 &0 & \{ \phi^{1}(x), \Psi_{b}(y) \}\\
\phi^{0}_{a}(x)    &0 &0&0 & 0& 0  & 0 &0 &0\\
\phi^{1}_{a}(x)   & 0 &\{ \phi^{1}_{a} (x), \phi^{b} (y) \}&0&0&0&0 &  \{ \phi^{1}_{a}(x), \Psi(y) \} & \{ \phi^{1}_{a}(x), \Psi_{b}(y) \} \\
\Psi(x) & \{ \Psi(x), \phi(y) \} &  \{ \Psi(x), \phi^{b}(y) \} &0 & 0&0&  \{ \Psi(x), \phi^{1}_{b}(y) \}&0 &0\\
\Psi_{a}(x) & \{ \Psi_{a}(x), \phi(y) \} &  \{ \Psi_{a}(x), \phi^{b}(y) \} &0 &  \{ \Psi_{a}(x), \phi^{1}(y) \}   &0&  \{ \Psi_{a}(x), \phi^{1}_{b}(y) \}&0 &0\\
\end{array} \right]. \nonumber \\
\end{eqnarray*}

This matrix has a vanishing determinant which tells us that there is at least one first class constraints. After a long calculation, we found that this matrix has a rank=6 and 6 null vectors, thus, the theory presents a set of 6 first class constraints and 6 second class constraints. The set of  null vectors of the  matrix (\ref{25}) are given by

\begin{eqnarray}
V^{1}&=&(-\epsilon{_{a}}^{c}B_{c}\delta(x-y),-\epsilon_{ba}B(1-2Bk^{2})\delta(x-y),0,-\frac{1}{l}\epsilon_{ab}e^{b}_{1}(1-4Bk^{2})\delta(x-y),0,\nonumber\\ &\,&l(\delta^{b}_{a}\partial_{1}\delta(x-y)-\epsilon{^{b}}_{a}\omega_{1}\delta(x-y)),\delta^{b}_{a}\delta(x-y)),\nonumber \\
V^{2}&=&(0,-\epsilon{^{c}}_{b}B_{c}\delta(x-y),0,\partial_{1}\delta(x-y),0,\epsilon{^{b}}_{c}e^{c}_{1}\delta(x-y),\delta(x-y),0), \nonumber \\
V^{3}&=&(0,0,\delta^{b}_{a}\delta(x-y),0,0,0,0,0), \nonumber \\
V^{4}&=&(0,0,0,0,\delta(x-y),0,0,0).
\label{25a}
 \end{eqnarray}


In order to identified the following 6 first class constraints, we used  the contraction of these null vectors with the constraints (\ref{18}) and (\ref{23})

\begin{eqnarray}
\gamma &=& \Psi+\partial_{1}\chi^{1}+\epsilon{^{a}}_{b}e^{b}_{1}\chi^{1}_{a}+\epsilon{_{a}}^{b}B_{b}\chi^{a}\approx 0, \nonumber \\
\gamma_{a}&=& \Psi_{a}+lD_{1}\chi^{1}_{a}-\frac{1}{l}\epsilon_{ab}e^{b}_{1}\chi^{1}(1-4k^{2}B)+\epsilon_{ab}(B\chi^{b}(1-2k^{2}B)-\chi B^{b})  \approx 0,\nonumber\\
\gamma^{0}_{a} &=& \phi^{0}_{a}=\Pi^{0}_{a}\approx 0,\nonumber\\
\gamma^{0}  &=&\phi^{0}=\Pi^{0}\approx 0,
\label{26}
\end{eqnarray}
and the following 6 second class constraints
\begin{eqnarray}
\chi        &=& \phi=\Pi \approx 0,  \nonumber \\
\chi^{a}    &=& \phi^{a}=\Pi^{a} \approx 0,  \nonumber \\
\chi^{1}    &=& \phi^{1}=\Pi^{1}-B \approx 0,  \nonumber \\
\chi^{1}_{a}    &=& \phi^{1}_{a}=\Pi^{1}_{a}-\frac{1}{l}B_{a} \approx 0.
\label{27}
\end{eqnarray}

It is worth noting that these constraints have not been previously documented in the literature. The comprehensive description of these constraints across the entire phase space is essential for understanding the fundamental gauge transformations and for establishing the Dirac brackets. Moreover, these constraints will prove pivotal in advancing the process of quantization.


After identifying and categorizing all constraints as first or second-class, it is crucial to analyze the Poisson algebra associated with these constraints. This analysis is essential to validate that the separation has been carried out accurately. Next, we will proceed to calculate the algebra of the constraints, expressing it as follows:

\begin{eqnarray}
\{\gamma (x),\gamma (y) \} &=& 0, \nonumber \\
\{\gamma_{a} (x),\gamma (y) \} &=& \epsilon{_{a}}^{b}\gamma_{b}\delta(x-y)\approx0, \nonumber \\
\{\gamma_{a} (x),\gamma_{b} (y) \} &=& -\epsilon_{ab}(1-4k^{2}B)\gamma\delta(x-y)+4k^{2}\epsilon_{ab}\chi^{1}\Psi\delta(x-y)\approx0, \nonumber \\
\{\gamma (x),\chi (y) \} &=& 0, \nonumber \\
\{\gamma (x),\chi^{b} (y) \} &=& \epsilon{_{a}}^{b}\chi^{a}\delta(x-y)\approx0, \nonumber \\
\{\gamma (x),\chi^{1} (y) \} &=& 0, \nonumber \\
\{\gamma (x),\chi^{1}_{b} (y) \} &=&-\epsilon{_{b}}^{a}\chi^{1}_{a}\approx 0, \nonumber \\
\{\gamma_{a} (x),\chi (y) \} &=& \frac{4}{l}k^{2}\epsilon_{ab}e^{b}_{1}\chi^{1}\delta(x-y)+\epsilon_{ab}(1-4k^{2}B)\chi^{b}\approx0, \nonumber \\
\{\gamma_{a} (x),\chi^{b} (y) \} &=& -\epsilon{_{a}}^{b}\chi\delta(x-y)\approx0, \nonumber \\
\{\gamma_{a} (x),\chi^{1} (y) \} &=&l\epsilon{_{a}}^{b}\chi^{1}_{b}\delta(x-y)\approx0, \nonumber \\
\{\gamma_{a} (x),\chi^{1}_{b} (y) \} &=&-\frac{1}{l}\epsilon_{ab}\chi^{1}(1-4Bk^{2})\delta(x-y)\approx0, \nonumber \\
\{\chi (x),\chi^{1} (y) \} &=& \delta(x-y), \nonumber \\
\{\chi^{a} (x),\chi^{1}_{b} (y) \} &=& \frac{1}{l}\delta^{a}_{b}\delta(x-y).
\label{28}
\end{eqnarray}




In the previous section, we noted that setting $k = 0$ removes us from quadratic gravity. However, as $k$ approaches zero, we move towards the $SO(3)$ algebra\footnote{By defining $T_{ab}=\epsilon_{ab}T$ and letting $T_{a2}$ be $P_{a}$, we can reformulate the algebra: $[T,T]=0$, $[T,P_{a}]=\epsilon_{a}^{b}P_{b}$, $[P_{a},P_{b}]=-\epsilon_{ab}T$.} \cite{Val}. Additionally, the results in equation (\ref{28}) correspond well with the classification in references \cite{Dirac1, Amorim, He}.

\subsection{Physical degrees of freedom}

With the correct identification of the constraints, we can carry out the counting of degrees of freedom in the following form: there are $18$ canonical variables $(e^{a}_{\mu},\Pi^{\mu}_{a},\omega_{\mu},\Pi^{\mu},B,\Pi,B_{a},\Pi^{a})$, $6$ first class constraints $(\gamma,\gamma_{a},\gamma^{0}_{a},\gamma^{0})$ and $6$ second class constraints $(\chi,\chi^{a},\chi^{1}_{a},\chi^{1})$ and one concludes that the  $S[e^{a}_{\mu},\omega_{\mu},B,B_{a}]$ action for gravity in two dimensions is devoid of degrees of freedom, therefore, the theory is topological. Certainly, when we treat the second class constraints given in (\ref{27}) as strong equations, we find that the aforementioned relationships simplify to the conventional constraints described in \cite{Val}.

\subsection{The extended action and Hamiltonian}


Moreover, the process of identifying Lagrange multipliers along with the constraints will enable us to establish the extended action. Through the utilization of the first-class constraints (\ref{26}), the second-class constraints (\ref{27}), and the Lagrange multipliers (\ref{24}), we determine that the extended action can be expressed in the following form

\begin{eqnarray}
S_{E}[Q_{L},P^{L}, \lambda_{A}, u_{A}]&=&\int d^{2}x (\dot{Q}_{L}P^{L}-\mathcal{H}-\tilde{\lambda}\gamma-\tilde{\lambda}^{a}\gamma_{a}-\check{\lambda}_{0}\gamma^{0}-\check{\lambda}^{a}_{0}\gamma^{0}_{a}-u\chi-u_{a}\chi^{a}\nonumber\\ &\,&-u^{a}_{1}\chi^{1}_{a}-u_{1}\chi^{1}),
\label{34}
\end{eqnarray}

and
\begin{eqnarray}
\mathcal{H}=-\omega_{0}\gamma-\frac{e^{a}_{0}}{l}\gamma_{a},
\label{35}
\end{eqnarray}

where $Q_{L}=(e^{a}_{\mu},\omega_{\mu}, B, B_{a})$, $P^{L}=(\Pi^{\mu}_{a}, \Pi^{\mu}, \Pi, \Pi^{a})$ and  $\lambda_{A}=(\tilde{\lambda},\tilde{\lambda}^{a},\check{\lambda}_{0},\check{\lambda}^{a}_{0})$, $u_{A}=(u,u_{a},u^{a}_{1},u_{1})$, are the Lagrange multipliers that enforce the first and second class constraints. We are to observable, by considering the second class constraints as strong equation, that the Hamiltonian (\ref{35}) is reduced to the usual expression defined on a reduced phase space context. From the extend action we can identify the extend Hamiltonian, which is given by

\begin{eqnarray}
H_{E}=\int dx (\mathcal{H}+\tilde{\lambda}\gamma+\tilde{\lambda}^{a}\gamma_{a}+\check{\lambda}^{a}_{0}\gamma^{0}_{a}+\check{\lambda}_{0}\gamma^{0}),
\label{36}
\end{eqnarray}

thus, the extended Hamiltonian is a linear combination of first-class constraints as expected.

\subsection{Dirac brackets}

Considering all the information we have gathered thus far, we are ready to formulate the Dirac brackets. There are two approaches to constructing them. The first method involves eliminating the second-class constraints and retaining the first-class ones, while the second approach entails fixing the gauge and transforming the first-class constraints into second-class ones. In this section, we will focus on the former approach, which involves eliminating the second-class constraints and preserving the first-class ones. To derive the Dirac brackets, we will utilize a matrix whose elements exclusively consist of Poisson brackets between the second-class constraints, denoted as $C^{\alpha\beta}(x, y) = \{\zeta^{\alpha}(x), \zeta^{\beta}(y)\}$. A straightforward yields

\begin{eqnarray}
[C^{\alpha\beta}(x,y)]=
\left(
 \begin{array}{cccc}
   0&0&\delta(x-y)&0\\
   0&0&0&\frac{1}{l}\delta^{a}_{b}\delta(x-y)\\
   -\delta(x-y)&0&0&0\\
   0&-\frac{1}{l}\delta^{b}_{a}\delta(x-y)&0&0\\
  \end{array}
\right),
\label{29}
\end{eqnarray}
its inverse is given by
\begin{eqnarray}
[C_{\alpha\beta}(x,y)]=
\left(
 \begin{array}{cccc}
   0&0&-\delta(x-y)&0\\
   0&0&0&-l\delta^{a}_{b}\delta(x-y)\\
   \delta(x-y)&0&0&0\\
   0&l\delta^{b}_{a}\delta(x-y)&0&0\\
  \end{array}
\right).
\label{30}
\end{eqnarray}
The Dirac  brackets among two functionals $A$, $B$ are expressed by
\begin{eqnarray}
\{A(x),B(y)\}_{D}=\{A(x),B(y)\}_{P}-\int dudv\{A(x),\zeta^{\alpha}(u)\}C_{\alpha\beta}(u,v)\{\zeta^{\beta}(v),B(y)\}.
\label{31}
\end{eqnarray}


The standard Poisson bracket, denoted as $\{A(x), B(y)\}_P$, is applied to the functionals $A$ and $B$. Meanwhile, the set of second class constraints, represented as $\zeta^{\alpha} = (\chi, \chi^a, \chi^1, \chi^1_a)$, is involved. Consequently, with the utilization of equations (\ref{30}) and (\ref{31}), we derive the Dirac brackets for the theory as follows

\begin{eqnarray}
\{e^{a}_{\mu}(x),\Pi^{\nu}_{b}(y)\}_{D}&=&\delta^{a}_{b}\delta^{\nu}_{\mu}\delta(x-y),\nonumber \\
\{e^{a}_{1}(x),B_{b}(y)\}_{D}&=& l\delta^{a}_{b}\delta(x-y),\nonumber \\
\{\omega_{1}(x),B(y)\}_{D}&=& \delta(x-y),\nonumber \\
\{\omega_{\mu}(x),\Pi^{\nu}(y)\}_{D}&=&\delta^{\nu}_{\mu}\delta(x-y).
\label{32}
\end{eqnarray}

It is noteworthy to mention that in the Polynomial BF action, the fields $(e^{a}_{1}, B_{a}, \omega_{1}, B)$ have acquired a non-commutative nature, just as it happens in the JT model \cite{Jaime1} with the fields $(A^{i}_{x}, \phi_{i})$.

By using the brackets given in (\ref{32}), we calculate the Dirac brackets among the first class and second class constraints given by

\begin{eqnarray}
\{\gamma(x),\gamma(y)\}_{D}&=& 0,\nonumber \\
\{\gamma_{a}(x),\gamma\}_{D}&=& \epsilon{_{a}}^{b}(\gamma_{b}+\epsilon_{bc}(1-2k^{2}B)\chi^{c}\chi^{1}+l\epsilon{^{c}}_{b}\chi^{1}_{c}\chi)\delta(x-y) , \nonumber \\
\{\gamma_{a} (x),\gamma_{b} (y) \}_{D} &=& -\epsilon_{ab}[(1-4k^{2}B)\gamma+\epsilon{^{c}}_{d}\chi^{1}_{c}(l(1-4k^2B)\chi^{d}+4k^{2}e^{d}_{1}\chi^{1})]\delta(x-y). \nonumber \\
\label{33}
\end{eqnarray}

We can observe that only squares of second class constraints appear. In fact, the Dirac brackets among first class constraints must be square of second class constraints and linear of first class constraints \cite{Dirac1, Amorim, He}.

\subsection{Gauge generator}


Currently, we are able to compute gauge transformations across the full phase space. Achieving the accurate gauge symmetry involves employing Dirac's conjecture to build a gauge generator utilizing the first-class constraints. The arrangement of constraints outlined in the full phase space will furnish us with the fundamental gauge transformations. To accomplish this, we will utilize Castellani's algorithm for the generation of the gauge generator.

We establish the gauge transformation generator as

\begin{eqnarray}
G=\int_{\sum}\left[D_{0}\tau^{a}\gamma^{0}_{a}+ D_{0}\tau\gamma^{0}+\theta\gamma+\theta^{a}\gamma_{a}\right].
\label{37}
\end{eqnarray}

Hence, we determine that the gauge transformations in the phase space are

\begin{eqnarray}
\delta{_{0}} B   &=& -\epsilon{_{ab}}\theta^{a}B^{b}, \nonumber\\
\delta{_{0}} B_{a}&=& \epsilon_{ab}\theta B^{b}-\epsilon_{ab}\theta^{b}B(1-2k^{2}B),\nonumber\\
 \delta{_{0}} \omega_{0}&=& D_{0}\tau, \nonumber \\
\delta{_{0}}\omega_{1}&=& -\partial_{1}\theta-\frac{1}{l}\epsilon_{ab}\theta^{a} e^{b}_{1}(1-4k^{2}B),\nonumber\\
\delta{_{0}} e^{a}_{0} &=& D_{0}\tau^{a}, \nonumber \\
\delta{_{0}} e^{a}_{1}&=&-lD_{1}\theta^{a}+\epsilon{^{a}}_{b}\theta e^{b}_{1},\nonumber \\
\delta{_{0}}\Pi&=& -\frac{4}{l}k^{2}\epsilon_{ab}\theta^{a}e^{b}_{1}\chi^{1}-\epsilon_{ab}(1-4k^{2}B)\theta^{a}\chi^{b},\nonumber \\
\delta{_{0}}\Pi^{a}&=& \epsilon{^{a}}_{b}\theta\chi^{b}-\epsilon_{ab}\theta^{b}\chi,\nonumber \\
\delta{_{0}}\Pi^{0}&=& 0, \nonumber \\
\delta{_{0}} \Pi^{1}&=& -\epsilon{_{ab}}\theta^{a}B^{b}-l\epsilon{_{a}}^{b}\theta^{a}\chi^{1}_{b}, \nonumber \\
\delta{_{0}}\Pi^{0}_{a}&=&0,\nonumber \\
\delta{_{0}}\Pi^{1}_{a}&=& \epsilon{_{a}}^{b}\theta\Pi^{1}_{b}-\frac{1}{l}\epsilon_{ab}\theta^{b}B(1-2k^{2}B)-\frac{1}{l}\epsilon_{ab}\theta^{b}\chi^{1}(1-4k^{2}B).
\label{38}
\end{eqnarray}


Although previously analyzed in the reduced phase space \cite{Val}, the gauge transformations in the full phase space (\ref{38}) have not been fully reported. Upon comparing our transformations with the earlier work, we observe a general agreement in the configuration variables, except for the transformation of the field $e^{a}_{1}$, where we identify an additional term $\epsilon{^{a}}_{b}\theta e^{b}_{1}$ not present in the reduced phase space analysis of \cite{Val}.


On the other hand, by redefining the parameter as follows: $\tau=-\theta=\zeta$, $\tau^{a}=-l\theta^{a}=\xi^{a}$, the gauge transformations (\ref{38}) can be expressed in a covariant form. This leads to the following gauge symmetry

\begin{eqnarray}
B &\rightarrow& B + \frac{1}{l}\epsilon_{ab}\xi^{a}B^{b}, \nonumber \\
B_{a} &\rightarrow& B_{a} -\epsilon{_{a}}^{b}\zeta B_{b} + \frac{1}{l}\epsilon_{ab}\xi^{b}B(1-2k^{2}B), \nonumber \\
e^{a}_{\mu} &\rightarrow& e^{a}_{\mu}+ D_{\mu}\xi^{a}-\epsilon{^{a}}_{b}\zeta e^{b}_{\mu}, \nonumber \\
\omega_{\mu} &\rightarrow& \omega_{\mu} + \partial_{\mu}\zeta + \frac{1}{l^{2}}\epsilon_{ab}(1-4k^{2}B)\xi^{a}e^{b}_{\mu}.
\label{39}
\end{eqnarray}

The results of (\ref{39}) are sufficient for the construction of Poincar\'e transformation. Starting from redefine the gauge parameters as $\zeta=-\tilde{\zeta}+\omega_{\mu}V^{\mu}$ and $\xi^{a}=e^{a}_{\mu}V^{\mu}$ \cite{Bla}

\begin{eqnarray}
\delta_{0} B      &=& \delta_{PGT}B-\epsilon_{\mu\nu}\frac{\delta S}{\delta \omega_{\mu}}V^{\nu},\nonumber \\
\delta_{0} B _{a} &=& \delta_{PGT} B_{a}-\varepsilon_{\mu\nu}\frac{\delta S}{\delta e^{a}_{\mu}}V^{\nu}, \nonumber \\
\delta_{0} e^{a}_{\mu}  &=& \delta_{PGT}e^{a}_{\mu}+\varepsilon_{\mu\nu}\frac{\delta S}{\delta B_{a}}V^{\nu}, \nonumber \\
\delta_{0} \omega_{\mu}&=&  \delta_{PGT}\omega_{\mu}+\varepsilon_{\mu\nu}\frac{\delta S}{\delta B }V^{\nu},
\label{42a}
\end{eqnarray}

where

\begin{eqnarray}
\delta_{PGT}B&=&V^{\mu}\partial_{\mu}B, \nonumber \\
\delta_{PGT}B_{a}&=&V^{\mu}\partial_{\mu}B_{a}+\epsilon{_{a}}^{b}B_{b}\tilde{\zeta},\nonumber \\
\delta_{PGT}e^{a}_{\mu}&=&e^{a}_{\alpha}\partial_{\mu}V^{\alpha}+\partial_{\alpha}e^{a}_{\mu}V^{\alpha}+\epsilon{^{a}}_{b}e^{b}_{\mu}\tilde{\zeta},\nonumber\\
\delta_{PGT}\omega_{\mu}&=&-\partial_{\mu}\tilde{\zeta}+\omega_{\alpha}\partial_{\mu}V^{\alpha}+\partial_{\alpha}\omega_{\mu}V^{\alpha}.
\label{43}
\end{eqnarray}

We can see that the gauge symmetries (\ref{42a}) take back to the Poincar\'e symmetries up to terms proportional to the equations of motion (\ref{10}).



The fundamental gauge transformations of the model are expressed by (\ref{39}) and do not align with diffeomorphisms. However, in any theory characterized by background independence, diffeomorphisms covariance is fundamental. This symmetry should emerge from the foundational gauge transformation. Hence, diffeomorphisms can be identified by redefining the gauge parameters as $\xi^{a}=e^{a}_{\mu}V^{\mu}$ and $\zeta=\omega_{\mu}V^{\mu}$, where $V$ represents a vector field

\begin{eqnarray}
\delta{_{0}} B &=& \frac{1}{l}\epsilon{_{a}}^{b}e^{a}_{\mu}V^{\mu}B_{b},\nonumber \\
\delta{_{0}} B_{a} &=& -\epsilon{_{a}}^{b}\omega_{\mu}V^{\mu}B_{b}+ \frac{1}{l}\epsilon_{ab}B(1-2k^{2}B)e^{b}_{\mu}V^{\mu}, \nonumber \\
\delta{_{0}} e^{a}_{\mu} &=& \partial_{\mu}(e^{a}_{\alpha}V^{\alpha})+\epsilon{^{a}}_{b}\omega_{\mu}(e^{b}_{\alpha}V^{\alpha})-\epsilon{^{a}}_{b}e^{b}_{\mu}\omega_{\alpha}V^{\alpha}, \nonumber \\
\delta{_{0}}\omega_{\mu}&=& \partial_{\mu}(\omega_{\alpha}V^{\alpha})+\frac{1}{l^{2}}\epsilon_{ab}(1-4k^{2}B)e^{a}_{\alpha}e^{b}_{\mu}V^{\alpha},
\label{40}
\end{eqnarray}

the gauge transformation (\ref{39}) is given by the following expression

\begin{eqnarray}
B &\rightarrow& B + \mathfrak{L}_{V}B-\epsilon_{\mu\alpha}\frac{\delta S}{\delta \omega_{\mu}}V^{\alpha}, \nonumber \\
B_{a} &\rightarrow& B_{a} + \mathfrak{L}_{V}B_{a}-\epsilon_{\mu\alpha}\frac{\delta S}{\delta e^{a}_{\mu}} V^{\alpha}, \nonumber\\
e^{a}_{\mu} &\rightarrow& e^{a}_{\mu} + \mathfrak{L}_{V}e^{a}_{\mu}+ \epsilon_{\mu\alpha}\frac{\delta S}{\delta B_{a}}V^{\alpha}, \nonumber\\
\omega_{\mu} &\rightarrow& \omega_{\mu}+ \mathfrak{L}_{V}\omega_{\mu}+\epsilon_{\mu\alpha}\frac{\delta S}{\delta B}V^{\alpha}.
\label{41}
\end{eqnarray}

Where the symbol $\mathfrak{L}$ represents the Lie derivative.
Therefore, diffeomorphisms are obtained as an internal symmetry of the theory from the fundamental gauge transformations (on shell).

In conclusion, we conducted a pure Hamiltonian analysis for the action (\ref{9}). This analysis yielded the extended action, extended Hamiltonian, the complete structure of constraints on the full phase space, their algebra, the count of degrees of freedom, and the fundamental gauge transformations. While working on the complete phase space introduces a set of first and second-class constraints, using the second-class constraints allows us to construct Dirac brackets, which will be valuable in the quantization of the theory. It is relevant to emphasize that when examining the scenario with k=0 in the model, results associated with the Jackiw-Teitelboim model are obtained.

\section{Faddeev-Jackiw Analysis }

In this section, we examine the BF Polynomial action within the framework of the modified Faddeev-Jackiw formalism \cite{Long, Liao}. Throughout this analysis, we not only address the constraint structure but also provide a detailed description of all the fundamental brackets present in the system. Furthermore, we carry out the determination of gauge transformations of the symplectic variables.

\subsection{Symplectic formalism}





An essential prerequisite for the Faddeev–Jackiw formalism is the presence of an initial Lagrangian of first order \cite{Fa}

\begin{eqnarray}
\mathcal{L}(\xi, \dot{\xi})= K^{(0)}_{i}\dot{\xi}^{(0)i}-V^{(0)}(\xi), \qquad i=1, ... , N.  
\label{43a}
\end{eqnarray}


Here, $\xi^{(0)i}$ represents a set of symplectic variables, with the canonical 1-form denoted by $\mathrm{K}^{(0)}_{i}$, and $V^{(0)}(\xi^{(0)})$ represents the initial symplectic potential.

Within the Faddeev-Jackiw (FJ) framework, the Euler-Lagrange equations of motion are \cite{Fa, Bar, Bar1}

\begin{equation}
\int f^{(0)}_{ij}(x,y)\dot{\xi}^{j}(y)dy=\frac{\delta}{\delta \xi^{i}(x)}\int V^{(0)}(y)dy.
\label{44a}
\end{equation}

Additionally, the construction of a symplectic matrix $f^{(0)}_{ij}$ is defined through the canonical 1-form $\mathrm{K}^{(0)}_{i}$

\begin{equation}
f^{(0)}_{ij}(x,y)=\frac{\delta \mathrm{K}_{j}(y)}{\delta\xi^{i}(x)}-\frac{\delta \mathrm{K}_{i}(x)}{\delta\xi^{j}(y)}.
\label{45a}
\end{equation}




If $f^{(0)}$ is regular, all symplectic variables can be determined using (\ref{44a}). However, if $f^{(0)}$ is singular, it implies the presence of constraints in the system. Suppose the rank of $f^{(0)}_{ij}$ is $r$, then $N-r= M$, In this scenario, there are $M$ zero-mode vectors $(v^{(0)}_{i})^{T}_{\alpha}$ of (\ref{45a}). These zero-mode vectors satisfy the equation

\begin{equation}
(v^{(0)}_{i})^{T}_{\alpha} f^{(0)}_{ij}(x,y)=0, \qquad \alpha=1, ... , M.
\label{45b}
\end{equation}


Hence, by utilizing the equation of motion (\ref{44a}), it is possible to formulate

\begin{eqnarray}
\Omega^{(0)}_{\alpha}&=&\int dx  (v^{(0)}_{i})^{T}_{\alpha}(x)\frac{\delta}{\delta\xi^{(0)i}(x)}\int dy V^{(0)}(\xi).
\label{45c}
\end{eqnarray}

In the FJ symplectic formalism, the quantities $\Omega^{(0)}_{\alpha}$ represent restrictions. 


From (\ref{13}), the first-order symplectic Lagrangian density of Polynomial BF action is given by 

\begin{eqnarray}
\mathcal{L}^{(0)}=B\dot{\omega}_{1}+\frac{1}{l}B_{a}\dot{e}^{a}_{1}- V^{(0)}, 
\label{43b}
\end{eqnarray}

here the superscript $^{(0)}$ means initial Lagrangian and where $V^{(0)}=-\omega_{0}(\partial_{1}B+\frac{1}{l}\epsilon{^{a}}_{b}B_{a}e^{b}_{1})-\frac{e^{a}_{0}}{l}(D_{1}B_{a}-\frac{1}{l}\epsilon_{ab}e^{b}_{1}B(1-2k^{2}B))$ is called the symplectic  potential.


In the [FJ] framework, it should be emphasized that the choice of symplectic variables is flexible, allowing us to opt for either the configuration variables or the phase space variables. Previous sections involved the construction of Dirac brackets through the elimination of second-class constraints. Consequently, in order to derive these outcomes within the [FJ] framework, we will opt for the configuration space as the symplectic variables \cite{Fa, Bar, Bar1}. To achieve this, we specifically select the following symplectic variables  $ \xi{^{(0)i}}(x)=\{B,B_{a},\omega_{0},\omega_{1},e^{a}_{0},e^{a}_{1}\}$ from the symplectic Lagrangian $\mathcal{L}^{(0)}$, and the components of the symplectic 1-forms are  $\mathrm{K}^{(0)}_{i}(x)=\{0,0,0,B,0,\frac{1}{l}B_{a}\}$. Therefore, employing our chosen set of symplectic variables results in the symplectic matrix (\ref{45a}) taking a specific form

\begin{table}[ht]
\begin{eqnarray}
f^{(0)}_{ij}(x,y) = \left[ 
\begin{array}{l|cccccc}
       & B  & B_{b} & \omega_{0} & \omega_{1} & e^{b}_{0}  & e^{b}_{1}     \\ \hline
B    & 0&0&0&1&0&0\\
B_{a}    &  0&0&0&0&0&\frac{1}{l}\delta^{a}_{b}\\
\omega_{0}    & 0&0&0&0&0&0\\
\omega_{1}    &  -1&0&0&0&0&0\\
e^{a}_{0}    &  0&0&0&0&0&0\\
e^{a}_{1}    &  0&-\frac{1}{l}\delta^{b}_{a}&0&0&0&0
\end{array}
\right]\delta(x-y).
\label{47a}
\end{eqnarray}
\end{table}

Clearly, the symplectic matrix (\ref{47a}) is singular and has dimensions of $[9\times9]$. Upon calculating the matrix rank, which is found to be 6, we deduce that M=9-6=3. This fact results in the presence of zero-modes $(v_{i}^{(0)})_{1}^{T}=(0,0,v^{\omega_{0}},0,0,0)$ and $(v_{i}^{(0)})_{2}^{T}=(0,0,0,0,v^{e^{a}_{0}},0)$, where $v^{\omega_{0}}$ and $v^{e^{a}_{0}}$ are arbitrary functions. Consequently, by using (\ref{45c}), the constraints we compute are

\begin{eqnarray}
\Omega^{(0)}&=&\int dx(v^{(0)}_{i})^{T}_{1}(x)\frac{\delta}{\delta\xi^{(0)i}(x)}\int dy V^{(0)}(\xi), \nonumber \\
            &=& -\int dx v^{\omega_{0}}(x)(\partial_{1}B+\frac{1}{l}\epsilon{^{a}}_{b}B_{a}e^{b}_{1}),\nonumber  \\
&\rightarrow& -(\partial_{1}B+\frac{1}{l}\epsilon{^{a}}_{b}B_{a}e^{b}_{1})=0,\nonumber  \\
\Omega^{(0)}_{a}&=&\int dx(v^{(0)}_{i})^{T}_{2}(x)\frac{\delta}{\delta\xi^{(0)i}(x)}\int dy V^{(0)}(\xi), \nonumber \\
            &=& -\int dx v^{e^{a}_{0}}(x)(D_{1}B_{a}-\frac{1}{l}\epsilon_{ab}e^{b}_{1}B(1-2k^{2}B)),\nonumber  \\
&\rightarrow& -(D_{1}B_{a}-\frac{1}{l}\epsilon_{ab}e^{b}_{1}B(1-2k^{2}B))=0,\nonumber  \\
    \label{44}
\end{eqnarray}

thus we identify the following constraints
\begin{eqnarray}
\Omega^{(0)}&=&-(\partial_{1}B+\frac{1}{l}\epsilon{^{a}}_{b}B_{a}e^{b}_{1})=0,\nonumber \\
\Omega^{(0)}_{a}&=&-(D_{1}B_{a}-\frac{1}{l}\epsilon_{ab}e^{b}_{1}B(1-2k^{2}B))=0,
\label{45}
\end{eqnarray}

the constraints acquired in (\ref{45}) are the secondary constraints identified through the Dirac method in the previous section. To investigate the presence of additional constraints, we will proceed with the implementation of the modified FJ formalism \cite{Long, Liao}. In the case where the symplectic matrix f demonstrates singularity, it becomes possible to convert the equations of motion (\ref{44a}) along with the consistency condition of the constraint (\ref{45c}), we have

\begin{eqnarray}
f^{(0)}_{ij}\dot{\xi}^{j}&=& \frac{\partial V}{\partial \xi^{i}}, \nonumber\\
\frac{\partial \Omega^{(0)}_{\alpha}}{\partial \xi^{i}}\dot{\xi}^{i}&=&0.
\label{45aa}
\end{eqnarray}

We can reformulate (\ref{45aa}) as

\begin{eqnarray}
   F^{(1)}_{ij}\dot{\xi}^{j}=Z_{i}(\xi).
\label{46}
\end{eqnarray}

At this stage, we have utilized the constraints identified (\ref{45c}) so far to derive the expression (\ref{46}), and then obtain

\begin{eqnarray}
F^{(1)}_{ij}= \left(
 \begin{array}{ccc}
   f^{(0)}_{ij}\\
   \frac{\partial\Omega^{(0)}}{\partial\xi^{j}}\\
   \frac{\partial\Omega^{(0)}_{a}}{\partial\xi^{j}}
  \end{array}
\right), \qquad    Z_{i}(\xi)=
\left(
 \begin{array}{cccc}
   \frac{\partial V^{(0)}(\xi)}{\partial \xi^{i}}\\
   0\\
   0\\
  \end{array}
\right).
\label{47}
\end{eqnarray}

Thus the symplectic matrix $F^{(1)}_{ij}$ is deduced as follows

\begin{table}[ht]
\begin{eqnarray}
F^{(1)}_{ij}(x,y) = \left(
 \begin{array}{ccc}
   f^{(0)}_{ij}\\
   \frac{\partial\Omega^{(0)}}{\partial\xi^{j}}\\
   \frac{\partial\Omega^{(0)}_{a}}{\partial\xi^{j}}
  \end{array}
\right)= \left[ 
\begin{array}{l|cccccc}
       & B  & B_{b} & \omega_{0} & \omega_{1} & e^{b}_{0}  & e^{b}_{1}     \\ \hline
B    & 0&0&0&1&0&0\\
B_{a}    &  0&0&0&0&0&\frac{1}{l}\delta^{a}_{b}\\
\omega_{0}    & 0&0&0&0&0&0\\
\omega_{1}    &  -1&0&0&0&0&0\\
e^{a}_{0}    &  0&0&0&0&0&0\\
e^{a}_{1}    &  0&-\frac{1}{l}\delta^{b}_{a}&0&0&0&0\\
\frac{\partial \Omega^{(0)}}{\partial \xi^{j}} & \frac{\partial \Omega^{(0)}}{\partial B}  & \frac{\partial \Omega^{(0)}}{\partial B_{b}}& \frac{\partial \Omega^{(0)}}{\partial \omega_{0}}& \frac{\partial \Omega^{(0)}}{\partial \omega_{1}}& \frac{\partial \Omega^{(0)}}{\partial e^{b}_{0}}& \frac{\partial \Omega^{(0)}}{\partial e^{b}_{1}} \\
\frac{\partial \Omega^{(0)}_{a}}{\partial \xi^{j}} & \frac{\partial \Omega^{(0)}_{a}}{\partial B}  & \frac{\partial \Omega^{(0)}_{a}}{\partial B_{b}}& \frac{\partial \Omega^{(0)}_{a}}{\partial \omega_{0}}& \frac{\partial \Omega^{(0)}_{a}}{\partial \omega_{1}}& \frac{\partial \Omega^{(0)}_{a}}{\partial e^{b}_{0}}& \frac{\partial \Omega^{(0)}_{a}}{\partial e^{b}_{1}} \\
\end{array}
\right]\delta(x-y).
\label{48}
\end{eqnarray}
\end{table}

The coefficient matrix (\ref{48}) mentioned in (\ref{46}) is non-square matrix.  However, despite this, it still exhibits linearly independent zero modes

\begin{eqnarray}
(v^{(1)}_{i})_{1}^{T}&=&(0,-\varepsilon{^{c}}_{a}B_{c}\delta(x-y),v^{\omega_{0}},\partial^{y}_{1}\delta(x-y),v^{e^{a}_{0}},\varepsilon{^{a}}_{c}e^{c}_{1}\delta(x-y),\delta(x-y),0),\nonumber \\
(v^{(1)}_{i})_{2}^{T}&=&(-\varepsilon{_{ab}}B^{b}\delta(x-y),\varepsilon_{ab}B(1-2k^{2}B)\delta(x-y),v^{\omega_{0}},-\frac{1}{l}\varepsilon_{ab}e^{b}_{1}(1-4k^{2}B)\delta(x-y),v^{e^{a}_{0}},\nonumber\\ &\,&l(\delta^{b}_{a}\partial^{y}_{1}\delta(x-y)+\varepsilon{_{a}}^{b}\omega_{1}\delta(x-y)),0,\delta^{b}_{a}\delta(x-y)).
\end{eqnarray}

Multiplying it on the left side of $F^{(1)}_{ij}$ results in zero. Expanding this process to both sides of equation (\ref{46}), we obtain the corresponding constraints \cite{Long, Liao}

\begin{equation}
(v^{(1)}_{i})_{c}^{T}Z_{i}=0,
\label{49}
\end{equation}

where $c=1,2$, we obtain that (\ref{49}) is an identity given by 

\begin{equation}
(v^{(1)}_{i})_{1}^{T}Z_{i}=-\frac{e^{a}_{0}}{l}\varepsilon{_{a}}{^{b}}\Omega^{(0)}_{b} \biggm|_{\Omega^{(0)}_{b}=0} =0,
\label{48a}
\end{equation}

and

\begin{equation}
(v^{(1)}_{i})_{2}^{T}Z_{i}=\omega_{0}\varepsilon{_{a}}^{b}\Omega^{(0)}_{b}\biggm|_{\Omega^{(0)}_{b}=0}+\frac{e^{b}_{0}}{l}\varepsilon_{ba}(1-4k^{2}B)\Omega^{(0)} \biggm|_{\Omega^{(0)}=0}=0,
\label{48b}
\end{equation}

thus, in [FJ]  formalism there are not more constraints for the theory under study. \\

Analogous to the conventional Faddeev–Jackiw approach, we have the ability to incorporate Lagrange multipliers $e^{a}_{0}=\dot{\lambda}^{a}$ and $ \omega_{0}=\dot{\lambda}$ that align with (\ref{48a}) and (\ref{48b}) into the Lagrangian, thereby formulating a novel Lagrangian 


\begin{equation}
\mathcal{L}^{(1)}=B\dot{\omega}_{1}+\frac{1}{l}B_{a}\dot{e}^{a}_{1}+\Omega^{(0)}_{a}\dot{\lambda}^{a}+\Omega^{(0)}\dot{\lambda}-V^{(1)}.
\label{50}
\end{equation}


In the scenario where $V^{(1)}=V^{(0)}\mid_{\Omega^{(0)}{a}=0,\Omega^{(0)}=0}=0$, the symplectic potential vanishes, highlighting the theory's general covariance, similar to the principle observed in General Relativity. Thus, referring to (\ref{50}), we can define a new set of symplectic variables $\xi{^{(1)i}}(x)=\{B,B_{a},\lambda,\omega_{1}, \lambda^{a},e^{a}_{1}\}$, accompanied by the corresponding symplectic 1-forms $\mathrm{K}^{(1)}_{i}(x)=\{0,0,\Omega^{(0)},B,\Omega^{(0)}_{a},\frac{1}{l}B_{a}\}$. By employing these revised symplectic variables and 1-forms, we can proceed to compute the symplectic matrix accordingly

\begin{table}[ht]
\begin{eqnarray*}
f^{(1)}_{ij}(x,y)=
\left[ 
\begin{array}{l|cccccc}
       & B  & B_{b} & \lambda & \omega_{1} & \lambda^{b} & e^{b}_{1}     \\ \hline
  B & 0&0& \partial^{y}_{1} & 1 &-\frac{1}{l}\epsilon_{bc}e^{c}_{1}(1-4k^{2}B)&0\\
  B_{a} & 0&0&\frac{1}{l}\epsilon{^{a}}_{c}e^{c}_{1}&0&\delta^{a}_{b}\partial^{y}_{1}-\epsilon{^{a}}_{b}\omega_{1}&\frac{1}{l}\delta^{a}_{b}\\
  \lambda & -\partial^{x}_{1}& -\frac{1}{l}\epsilon{^{b}}_{c}e^{c}_{1}&0&0&0&-\frac{1}{l}\epsilon{^{c}}_{b}B_{c}\\
 \omega_{1} & -1&0&0&0&\epsilon{_{b}}^{c}B_{c}&0\\
 \lambda^{a} & \frac{1}{l}\epsilon_{ab}e^{b}_{1}(1-4k^{2}B)&-\delta^{a}_{b}\partial^{x}_{1}-\epsilon{_{b}}^{a}\omega_{1}&0&-\epsilon{_{a}}^{c}B_{c}&0&\frac{1}{l}\epsilon_{ab}B(1-2k^{2}B)\\
 e^{a}_{1} & 0&-\frac{1}{l}\delta^{b}_{a}&\frac{1}{l}\epsilon{^{c}}_{a}B_{c}&0&\frac{1}{l}\epsilon_{ab}B(1-2k^{2}B)&0\\
  \end{array}
\right]\nonumber \\
\label{51}
\end{eqnarray*}
\end{table}

\begin{eqnarray}
\times \delta(x-y).
\label{51a}
\end{eqnarray}


\subsection{FJ brackets}

The symplectic matrix $f^{(1)}_{ij}$ is a singular $[9\times9]$ matrix. Despite this singularity, we have established that there are no additional constraints. The noninvertibility of (\ref{51}) signifies the existence of gauge symmetry within the theory. To eliminate redundant degrees of freedom and derive the desired Faddeev–Jackiw generalized brackets, additional gauge-fixing conditions must be applied. Since there are two linearly independent null eigenvectors, we opt for the following gauge constraints

\begin{eqnarray}
\omega_{0}&=&0,\nonumber\\
e^{a}_{0}(x)&=&0.
\label{52}
\end{eqnarray}




Following that, we add the Lagrangian multipliers $\alpha_a$ and $\theta$ corresponding to the previously discussed gauge fixing, with the goal of developing a new symplectic Lagrangian. The expression for the symplectic Lagrangian is now

\begin{equation}
\mathcal{L}^{(2)}=B\dot{\omega}_{1}+\frac{1}{l}B_{a}\dot{e}^{a}_{1}+(\alpha_{a} +\Omega^{(0)}_{a})\dot{\lambda}^{a}+(\theta+\Omega^{(0)})\dot{\lambda},
\label{53}
\end{equation}


therefore, we can characterize the following set of symplectic variables $\xi{^{(2)i}}(x)=\{B,B_{a},\lambda,\theta,\omega_{1}, \lambda^{a},\alpha_{a},e^{a}_{1}\}$ and the symplectic 1-forms $\mathrm{K}^{(2)}_{i}(x)=\{0,0,\Omega^{(0)}+\theta,0,B,\Omega^{(0)}_{a}+\alpha_{a},0,\frac{1}{l}B_{a}\}$. Additionally, by employing these symplectic variables, we establish the expression for the symplectic matrix

\begin{eqnarray*}
f^{(2)}_{ij}(x,y)=
\end{eqnarray*}

\begin{table}[ht]
\begin{eqnarray*}
\left[ 
\begin{array}{l|cccccccc}
       & B  & B_{b} & \lambda &\theta &\omega_{1} & \lambda^{b} & \alpha_{b} & e^{b}_{1}     \\ \hline
B &  0&0& \partial^{y}_{1} & 0 & 1 &-\frac{1}{l}\epsilon_{bc}e^{c}_{1}(1-4k^{2}B)&0&0\\
B_{a} &  0&0&\frac{1}{l}\epsilon{^{a}}_{c}e^{c}_{1}&0&0&\delta^{a}_{b}\partial^{y}_{1}-\epsilon{^{a}}_{b}\omega_{1}&0&\frac{1}{l}\delta^{a}_{b}\\
 \lambda & -\partial^{x}_{1}&-\frac{1}{l}\epsilon{^{b}}_{c}e^{c}_{1}&0&-1&0&0&0&-\frac{1}{l}\epsilon{^{c}}_{b}B_{c}\\
\theta &  0&0&1&0&0&0&0&0 \\
\omega_{1} &  -1&0&0&0&0&\epsilon{_{b}}^{c}B_{c}&0&0\\
 \lambda^{a} &  \frac{1}{l}\epsilon_{ab}e^{b}_{1}(1-4k^{2}B)&-\delta^{b}_{a}\partial^{x}_{1}-\epsilon{_{a}}^{b}\omega_{1}&0&0&-\epsilon{_{a}}^{c}B_{c}&0&-\delta^{b}_{a}&\frac{1}{l}\epsilon_{ab}B(1-2k^{2}B)\\
 \alpha_{a} & 0&0&0&0&0&\delta^{b}_{a}&0&0\\
 e^{a}_{1} & 0&-\frac{1}{l}\delta^{b}_{a}&\frac{1}{l}\epsilon{^{c}}_{a}B_{c}&0&0& \frac{1}{l}\epsilon_{ab}B(1-2k^{2}B)&0&0\\
  \end{array}
\right]\nonumber \\
\label{54}
\end{eqnarray*}
\end{table}

\begin{eqnarray}
\times \delta(x-y).
\label{54a}
\end{eqnarray}


Thus far, a non-singular matrix has been obtained, allowing for the deduction of its inverse matrix

\begin{eqnarray*}
[f^{(2)}_{ij}(x,y)]^{-1}=
\end{eqnarray*}

\begin{eqnarray*}
\left[ 
\begin{array}{l|cccccccc}
       & B  & B_{b} & \lambda &\theta &\omega_{1} & \lambda^{b} & \alpha_{b} & e^{b}_{1}     \\ \hline
 B &  0&0&0&0&-1&0&-\epsilon{_{a}}^{c}B_{c}&0\\
B_{a} &   0&0&0&\epsilon{^{c}}_{b}B_{c}&0&0&\epsilon_{ba}B(1-2k^{2}B)&-l\delta^{a}_{b}\\
\lambda &   0&0&0&1&0&0&0&0\\
\theta & 0&-\epsilon{^{c}}_{a}B_{c}&-1&0&\partial^{x}_{1}&0&0&\epsilon{^{a}}_{c}e^{c}_{1}\\
\omega_{1} & 1&0&0&-\partial^{y}_{1}&0&0&\frac{1}{l}\epsilon_{ac}e^{c}_{1}(1-4k^{2}B)&0\\
\lambda^{a} &0&0&0&0&0&0&\delta^{b}_{a}&0\\
\alpha_{a} & \epsilon{_{b}}^{c}B_{c}&\epsilon_{ba}B(1-2k^{2}B)&0&0&-\frac{1}{l}\epsilon_{bc}e^{c}_{1}(1-4k^{2}B)&-\delta^{a}_{b}&0&
l(\delta^{a}_{b}\partial^{x}_{1}+\varepsilon{_{b}}^{a}\omega_{1})\\
e^{a}_{1} & 0&l\delta^{b}_{a}&0&-\epsilon{^{b}}_{c}e^{c}_{1}&0&0& -l(\delta^{b}_{a}\partial^{y}_{1}+\varepsilon{^{b}}_{a}\omega_{1})&0 \\
  \end{array}
\right]
\label{55}
\end{eqnarray*}

\begin{eqnarray}
\times \delta(x-y).
\end{eqnarray}

By calculating the inverse of this symplectic matrix(\ref{55}), we can recognize the standard Faddeev–Jackiw generalized brackets for the physical variables through a defined procedure

\begin{eqnarray}
\{\xi_{i}^{(2)}(x),\xi_{j}^{(2)}(y)\}_{FJ}=[f^{(2)}_{ij}(x,y)]^{-1},
\end{eqnarray}
thus, the following brackets are identified
\begin{eqnarray}
\{\omega_{1}(x),B(y)\}_{FJ}&=&\delta(x-y),\nonumber \\
\{e^{a}_{1}(x),B_{b}(y)\}_{FJ}&=&l\delta^{a}_{b}\delta(x-y).
\label{56}
\end{eqnarray}



It is noteworthy that the generalized (FJ) brackets correspond to those derived using the Dirac method mentioned earlier. Specifically, a redefinition of the fields, introducing the momenta as provided by

\begin{eqnarray}
\Pi&=&0,  \nonumber \\
\Pi^{a}&=& 0,  \nonumber \\
\Pi^{1}_{a}&=&\frac{1}{l}B_{a},  \nonumber \\
\Pi^{1}&=&B.
\label{57}
\end{eqnarray}



It allows us to derive the following generalized Faddeev–Jackiw (FJ) brackets and demonstrate their equivalence with the Dirac brackets obtained earlier (\ref{32})

\begin{eqnarray}
\{e^{a}_{1}(x),\Pi^{1}_{b}(y)\}_{FJ}&=&\delta^{a}_{b}\delta(x-y),\nonumber \\
\{e^{a}_{1}(x),B_{a}(y)\}_{FJ}&=&l\delta^{a}_{b}\delta(x-y),\nonumber \\
\{\omega_{1}(x),\Pi^{1}(y)\}_{FJ}&=&\delta(x-y),\nonumber \\
\{\omega_{1}(x),B (y)\}_{FJ}&=&\delta(x-y).
\label{58}
\end{eqnarray}

Now, it is essential to introduce a bracket for observables on $\sum$. This operator must align with the commutator in the classical limit. In this context, for any pair of observables $\mathcal{O}_{1}$, $\mathcal{O}_{2}$ defined in the phase space, possessing a symplectic structure such as $\{\xi ^{final}_{i} , \xi^{final}_{j}\}$ , we can denote the following relationship \cite{Omar}

\begin{eqnarray}
\{ \mathcal{O}_{1}(\xi) , \mathcal{O}_{2}(\xi) \}_{FJ}= \sum_{ij}\int d^{3}r\frac{\delta \mathcal{O}_{1}(\xi)}{\delta \xi_{i}(r)}(f^{(final)}_{ij})^{-1} \frac{\delta \mathcal{O}_{2}(\xi)}{\delta \xi_{j}(r)}.
\end{eqnarray}

The transition of the Faddeev-Jackiw bracket by operator commutattion relations according to 

\begin{eqnarray}
\{ \mathcal{O}_{1}(\xi) , \mathcal{O}_{2}(\xi) \}_{FJ} \quad \rightarrow \frac{1}{i\hbar}[\mathcal{O}_{1}(\xi) , \mathcal{O}_{2}(\xi)].
\end{eqnarray}






\subsection{Path integral}

In this section, we will determine the transition amplitude \cite{Lliao, Toms}. To accomplish this, we will rewrite the matrix (\ref{54a}) as follows in order


\begin{table}[ht]
\begin{eqnarray*}
\left[ 
\begin{array}{l|cccccccc}
       & B  & B_{b} & \omega_{1} & e^{b}_{1} & \lambda &\theta & \lambda^{b} & \alpha_{b} \\ \hline
B &  0 & 0 & 1 & 0 &\partial^{y}_{1} & 0 -&\frac{1}{l}\epsilon_{bc}e^{c}_{1}(1-4k^{2}B)&0\\
B_{a} &  0&0&  0& \frac{1}{l}\delta^{a}_{b} & \frac{1}{l}\epsilon{^{a}}_{c}e^{c}_{1}&0&\delta^{a}_{b}\partial^{y}_{1}-\epsilon{^{a}}_{b}\omega_{1}&0\\
\omega_{1} &  -1&0&0&0&0& 0&\epsilon{_{b}}^{c}B_{c}&0\\
 e^{a}_{1} & 0&-\frac{1}{l}\delta^{b}_{a}&0&0&\frac{1}{l}\epsilon{^{c}}_{a}B_{c}&0&\frac{1}{l}\epsilon_{ab}B(1-2k^{2}B)&0\\
\lambda & -\partial^{x}_{1}&- \frac{1}{l}\epsilon{^{b}}_{c}e^{c}_{1}&0&- \frac{1}{l}\epsilon{^{c}}_{b}B_{c} &0 &-1&0&0\\
\theta &  0&0&0&0&1&0&0&0 \\
 \lambda^{a} &  \frac{1}{l}\epsilon_{ab}e^{b}_{1}(1-4k^{2}B)&-\delta^{b}_{a}\partial^{x}_{1}-\epsilon{_{a}}^{b}\omega_{1}&-\epsilon{_{a}}^{c}B_{c}&\frac{1}{l}\epsilon_{ab}B(1-2k^{2}B)& 0& 0& 0&-\delta^{b}_{a}\\
 \alpha_{a} & 0&0&0&0&0&0&\delta^{b}_{a}&0\\
  \end{array}
\right]\nonumber \\
\label{58a}
\end{eqnarray*}
\end{table}

\begin{eqnarray}
\times \delta(x-y).
\end{eqnarray}

To calculate the determinant of the symplectic two-form in (\ref{58a}), we will examine its general block form structure

\begin{eqnarray}
f^{(2)}_{ij}=
\left(
 \begin{array}{cc}
   \mathcal{A}&\mathcal{B}\\
   \mathcal{C}&\mathcal{D}\\
  \end{array}
\right), \qquad with \qquad \mathcal{C}=-\mathcal{B}^{T}.
\label{58aa}
\end{eqnarray}

In this context, the specific expression for each sub-matrix $\mathcal{A}$, $\mathcal{B}$, and $\mathcal{D}$ in equation (\ref{58aa}) is revealed to be

\begin{table}[ht]
\begin{eqnarray*}
\small{
\mathcal{A}=\left[ 
\begin{array}{l|cccc}
       & B  & B_{b} & \omega_{1} & e^{b}_{1} \\ \hline
B &  0 & 0 & 1 & 0\\
B_{a} &  0&0&  0& \frac{1}{l}\delta^{a}_{b}\\
\omega_{1} &  -1&0&0&0\\
 e^{a}_{1} & 0&-\frac{1}{l}\delta^{b}_{a}&0&0\\
  \end{array}
\right],  \quad  \mathcal{B}=\left[ 
\begin{array}{l|cccc}
       & \lambda &\theta & \lambda^{b} & \alpha_{b} \\ \hline
B & \partial^{y}_{1} & 0 &-\frac{1}{l}\epsilon_{bc}e^{c}_{1}(1-4k^{2}B)&0\\
B_{a} & \frac{1}{l}\epsilon{^{a}}_{c}e^{c}_{1}&0&\delta^{a}_{b}\partial^{y}_{1}-\epsilon{^{a}}_{b}\omega_{1}&0\\
\omega_{1} & 0& 0&\epsilon{_{b}}^{c}B_{c}&0\\
 e^{a}_{1} &\frac{1}{l}\epsilon{^{c}}_{a}B_{c}&0&\frac{1}{l}\epsilon_{ab}B(1-2k^{2}B)&0\\
  \end{array}
\right], \quad \mathcal{D}=\left[ 
\begin{array}{l|cccc}
       & \lambda &\theta & \lambda^{b} & \alpha_{b} \\ \hline
\lambda & 0 &-1&0&0\\
\theta  & 1&0&0&0 \\
 \lambda^{a} & 0& 0& 0&-\delta^{b}_{a}\\
 \alpha_{a}  & 0&0&\delta^{b}_{a}&0\\
  \end{array}
\right] }.\nonumber \\ 
\label{58bb}
\end{eqnarray*}
\end{table}

Using the standard identity for any matrix

\begin{eqnarray}
\left(
 \begin{array}{cc}
   \mathcal{A}&\mathcal{B}\\
   \mathcal{C}&\mathcal{D}\\
  \end{array}
\right) = \left(
 \begin{array}{cc}
   \mathcal{A}&0\\
   \mathcal{C}&\mathcal{I}\\
  \end{array}
\right) \left(
 \begin{array}{cc}
   \mathcal{I}&\mathcal{A}^{-1}\mathcal{B}\\
   0&  \mathcal{D}-\mathcal{C}\mathcal{A}^{-1}  \mathcal{B}\\
  \end{array}
\right),
\label{58cc}
\end{eqnarray}

assuming that $\mathcal{A}^{-1}$ exists, to see that

\begin{eqnarray}
det \left(
 \begin{array}{cc}
   \mathcal{A}&\mathcal{B}\\
   \mathcal{C}&\mathcal{D}\\
  \end{array}
\right) = (det  \mathcal{A})[det( \mathcal{D}- \mathcal{C} \mathcal{A}^{-1} \mathcal{B})],
\label{58dd}
\end{eqnarray}

Therefore, utilizing equation (\ref{58dd}) and performing algebraic manipulations, it becomes feasible to demonstrate the determinant of $f^{(2)}_{ij}$ is evaluated in terms of the symplectic two-form from and is equal

\begin{eqnarray}
det [f^{(2)}_{ij}]= \frac{1}{l^{4}}.
\label{58f}
\end{eqnarray}

Presently, it is noteworthy to observe that according to \cite{Lliao, Toms}, the functional measure within the path integral linked to our model, under the time gauge, is

\begin{eqnarray}
d\mu = (\prod_{i}[D\xi^{(2)i}])(det [f^{(2)}_{ij}(x,y)])^{1/2}=(\prod_{i}[D\xi^{(2)i}])[\frac{1}{l^{2}}]= (\prod_{i}[D\xi^{(2)i}]) \Lambda.
\label{58g}
\end{eqnarray}

It is evident that the path integral measure in (\ref{58g}) is associated with the cosmological constant. The approximation proposed by Batalin, Fradkin, and Vilkovisky (BFV) is utilized in \cite{Val}, incorporating the concept of constraints in Path Integral Quantization during the calculation of the transition amplitude. It is observed that the functional measure in the time gauge equals 1, a distinction from the measure identified in the current study.


The partition function incorporates the symplectic two-form straightforwardly; since this element is crucial for the Faddeev-Jackiw method, no additional effort is needed, except for calculating its determinant. In accordance with (\ref{58g}) and the Lagrangian (\ref{53}), the partition function takes the following form

\begin{eqnarray}
Z&=&\int d\mu exp[\frac{i}{\hbar}\int d^{3}x \mathcal{L}^{(2)}], \nonumber \\
&=&\int DBDB_{a}D\omega_{1}De^{a}_{1}D\lambda D\theta D\lambda^{a} D\alpha_{a}[\frac{1}{l^{2}}]exp[\frac{i}{\hbar}\int d^{3}x (B\dot{\omega}_{1}+\frac{1}{l}B_{a}\dot{e}^{a}_{1}+(\alpha_{a}+\Omega^{(0)}_{a})\dot{\lambda}^{a}+(\theta+\Omega^{(0)})\dot{\lambda})].
\end{eqnarray}



\subsection{FJ physical degrees of freedom}

As mentioned earlier, in the Faddeev–Jackiw (FJ) approach, there's no need to categorize constraints as first or second class, as all constraints hold the same significance. Consequently, the count of physical degrees of freedom can be performed as follows: with $6$ dynamical variables represented by 
$(B,\omega_{1},B_{a},e^{a}_{1},)$,  and $6$ constraints $(\Omega^{(0)}_{a},\Omega^{(0)},e^{a}_{0},\omega_{0})$, the theory exhibits no physical degrees of freedom.


\subsection{FJ gauge generator}

We finish this section by calculating the gauge transformations of the theory, for this aim we calculate the modes of the matrix (\ref{51a}), this mode are given by
\begin{eqnarray}
(w^{(1)}_{i})_{1}^{T}&=&(0,\epsilon{_{b}}^{c}B_{c}\delta(x-y),-\delta(x-y),\partial^{y}_{1}\delta(x-y),0,\epsilon{^{b}}_{c}e^{c}_{1}\delta(x-y)),\nonumber\\
(w^{(1)}_{i})_{2}^{T}&=&(-\epsilon{_{a}}^{c}B_{c}\delta(x-y),-\epsilon_{ab}B(1-2k^{2}B)\delta(x-y),0,-\frac{1}{l}\epsilon_{ac}e^{c}_{1}(1-4k^{2}B)\delta(x-y),-\delta^{b}_{a}\delta(x-y),l(\delta^{b}_{a}\partial^{y}_{1}\delta(x-y)\nonumber\\ &\,&+\epsilon{_{b}}^{a}\omega_{1}\delta(x-y))).
\label{59}
\end{eqnarray}


Consistent with the [FJ] symplectic formalism, the zero modes $(w^{(1)}_{i})_{1}^{T}$ and $(w^{(1)}_{i})_{2}^{T}$ act as the generator of the infinitesimal gauge transformation of the action (\ref{43b}). Specificall we have that

\begin{eqnarray}
\delta_{G}\xi^{(1)i}= \int dx (w^{(1)}_{i})^{T}_{c} \varepsilon^{c}
\end{eqnarray}

where $"\varepsilon^{c} "$ is a set of infinitesimal arbitrary parameters.

From the above, we can see that the gauge transformation is therefore given by \footnote{Where $\epsilon^{1}=\tilde{\theta}=-\tau$ and  $\epsilon^{2}=\tilde{\theta}^{a}=-\tau^{a}$.}
\begin{eqnarray}
\delta_{G} \xi^{(1)a}= (\delta_{G} B, \delta_{G} B_{a}, \delta_{G} \lambda, \delta_{G} \omega_{1}, \delta_{G} \lambda^{a}, \delta_{G} e^{a}_{1} )= \int dx [(w^{(1)}_{i})^{T}_{1} \varepsilon^{1}+ (w^{(2)}_{i})^{T}_{1} \varepsilon^{2}]   ,
\end{eqnarray}

 we obtain \footnote{coming back to the original variables, $\partial_{t}\delta{_{G}}\lambda=\delta_{G}\omega_{0}$ and $\partial_{t}\delta{_{G}}\lambda^{a}=\delta_{G}e^{a}_{0}$.}


\begin{eqnarray}
\delta{_{G}} B   &=& -\epsilon_{ab}\tilde{\theta}^{a}B^{b}, \nonumber\\
\delta{_{G}} B_{a}&=& \epsilon_{ab}\tilde{\theta}B^{b}-\epsilon_{ab}\tilde{\theta}^{b}B (1-2k^{2}B),\nonumber\\
\delta{_{G}} \omega_{0}&=& -\partial_{0} \tilde{\theta}= \partial_{0}\tau, \nonumber \\
\delta{_{G}}\omega_{1}&=& -\partial_{1}\tilde{\theta}-\frac{1}{l}\epsilon_{ab}\tilde{\theta}^{a}e^{b}_{1}(1-4k^{2}B), \nonumber \\
\delta{_{G}} e^{a}_{0} &=& -\partial_{0}\tilde{\theta^{a}}=\partial_{0}\tau^{a}, \nonumber \\
\delta{_{G}} e^{a}_{1}&=& -lD_{1}\tilde{\theta}^{a} + \epsilon{^{a}}_{b}\tilde{\theta}e^{b}_{1}.
\label{60}
\end{eqnarray}

Using the FJ symplectic framework, we have successfully reproduced most components of the gauge transformations described in the Dirac approach. However, a discrepancy is observed in the transformations of the fields $e^{a}_{0}$ between the FJ and Dirac approaches. Montani and Wotzasek (MW) \cite{Mon} propose a method to obtain complete transformations across the entire configuration space.
According to the MW method, one should write the functional variation of the corresponding Lagrangians\footnote{In term of matrix form $(\frac{\delta \mathcal{L}^{(1)}}{\delta \xi^{(1)i}})^{T}=
\left(
\begin{array}{cccccc}
\frac{\delta \mathcal{L}^{(1)}}{\delta \xi^{(1)1}} & \frac{\delta \mathcal{L}^{(1)}}{\delta \xi^{(1)2}} & \frac{\delta \mathcal{L}^{(1)}}{\delta \xi^{(1)3}} &
\frac{\delta \mathcal{L}^{(1)}}{\delta \xi^{(1)4}}&
\frac{\delta \mathcal{L}^{(1)}}{\delta \xi^{(1)5}}&
\frac{\delta \mathcal{L}^{(1)}}{\delta \xi^{(1)6}}
\end{array} 
\right)$.} to zero.  Therefore,

\begin{eqnarray}
\frac{\delta \mathcal{L}^{(1)}}{\delta \xi^{(1)i}}= f^{(1)}_{ij}\dot{\xi}^{(1)j}-\frac{\partial V^{(1)}}{\partial \xi^{(1)i}}. 
\label{59a}
\end{eqnarray}
So, according to MW, on multiplying (\ref{59a}) by the zero modes (\ref{59}), we have

\begin{eqnarray}
\epsilon^{c}(w^{(1)}_{i})^{T}_{c}\left(\frac{\delta \mathcal{L}^{(1)}}{\delta \xi^{(1)i}}= f^{(1)}_{ij}\dot{\xi}^{(1)j}-\frac{\partial V^{(1)}}{\partial \xi^{(1)i}}\right), \quad c=1,2.
\label{60}
\end{eqnarray}



from (\ref{60}), one finds the GT over all the configuration space. Using this equation, the GT  can be written as  

\begin{eqnarray}
(\epsilon^{1}(w^{(1)}_{i})^{T}_{1} + \epsilon^{2}(w^{(1)}_{i})^{T}_{2})\frac{\delta \mathcal{L}^{(1)}}{\delta \xi^{(1)i}}= (\epsilon^{1}(w^{(1)}_{i})^{T}_{1} + \epsilon^{2}(w^{(1)}_{i})^{T}_{2})f^{(1)}_{ij}\dot{\xi}^{(1)j}
\end{eqnarray}

Specifically we have that 

\begin{eqnarray}
\mathcal{A}_{1}\frac{\delta \mathcal{L}^{(1)}}{\delta B}+\mathcal{A}_{2}\frac{\delta \mathcal{L}^{(1)}}{\delta B_{a}}+\mathcal{A}_{3}\frac{\delta \mathcal{L}^{(1)}}{\delta \lambda}+\mathcal{A}_{4} \frac{\delta \mathcal{L}^{(1)}}{\delta \omega_{1}}+\mathcal{A}_{5}\frac{\delta \mathcal{L}^{(1)}}{\delta \lambda^{a}}+\mathcal{A}_{6}\frac{\delta \mathcal{L}^{(1)}}{\delta e^{a}_{1}}&=& \mathcal{A}_{1} (\dot{\omega}_{1}-\frac{1}{l}\epsilon_{cb}e^{b}_{1}(1-4k^{2}B)\dot{\lambda}^{c}-\partial_{1}\dot{\lambda})+\nonumber \\ 
 &&\mathcal{A}_{2}(\frac{1}{l}\dot{e}^{a}_{1}-\partial_{1}\dot{\lambda}^{a}+\epsilon{_{b}}^{a}\omega_{1}\dot{\lambda}^{b}+\frac{1}{l}\epsilon{^{a}}_{b}e^{b}_{1}\dot{\lambda})+\nonumber \\ 
 &&\mathcal{A}_{3}(-\partial_{t}\Omega^{(0)})+\mathcal{A}_{4}(-\dot{B}+\epsilon{_{a}}^{b}B_{b}\dot{\lambda}^{a})+\nonumber \\ 
 &&\mathcal{A}_{5}(-\partial_{t}\Omega^{(0)}_{a})+\mathcal{A}_{6}(-\frac{1}{l}\dot{B}_{a}-\frac{1}{l}\epsilon_{ba}B(1-2k^{2}B)\dot{\lambda}^{b}\nonumber \\ 
 &&+\frac{1}{l}\epsilon{^{b}}_{a}B_{b}\dot{\lambda})). \nonumber 
\end{eqnarray}

where $\mathcal{A}_{1}=-\epsilon{_{a}}^{c}B_{c}\epsilon^{2}, \mathcal{A}_{2}=\epsilon{_{a}}^{b}B_{b}\epsilon^{1}-\epsilon_{ab}B(1-2k^{2}B)\epsilon^{2}, \mathcal{A}_{3}=-\epsilon^{1},\mathcal{A}_{4}=[-\partial_{1} \epsilon^{1}-\frac{1}{l}\epsilon_{ab}e^{b}_{1}(1-4k^{2}B)\epsilon^{2}],\mathcal{A}_{5}=-\epsilon^{2},\mathcal{A}_{6}= \epsilon{^{a}}_{b}e^{b}_{1}\epsilon^{1}-lD_{1}\epsilon^{2}$.

Now, coming back to the original variables

\begin{eqnarray}
\dot{\lambda}&\rightarrow& \omega_{0},\nonumber \\
\dot{\lambda}^{a}&\rightarrow& e^{a}_{0}, \nonumber \\
\frac{\delta \mathcal{L}^{(1)}}{\delta \lambda}&\rightarrow& \frac{\delta \mathcal{L}^{(1)}}{\delta \omega_{0}}\partial_{t}, \nonumber \\
\frac{\delta \mathcal{L}^{(1)}}{\delta \lambda^{a}}&\rightarrow& \frac{\delta \mathcal{L}^{(1)}}{\delta e^{a}_{0}}\partial_{t}, \nonumber \\
\frac{\delta \mathcal{L}^{(1)}}{\delta B}&=& \frac{\delta \mathcal{L}^{(0)}}{\delta B},\nonumber \\
\frac{\delta \mathcal{L}^{(1)}}{\delta B_{a}}&=& \frac{\delta \mathcal{L}^{(0)}}{\delta B_{a}},\nonumber \\
\frac{\delta \mathcal{L}^{(1)}}{\delta \omega_{1}}&=& \frac{\delta \mathcal{L}^{(0)}}{\delta \omega_{1}},\nonumber \\
\frac{\delta \mathcal{L}^{(1)}}{\delta e^{a}_{1}}&=&\frac{\delta \mathcal{L}^{(0)}}{\delta e^{a}_{1}},
\label{60aa}
\end{eqnarray}

and implementing the equations of motion for the gauge fields

\begin{eqnarray}
\frac{\delta \mathcal{L}^{(0)}}{\delta \omega_{0}}&=&\Omega^{(0)}, \nonumber \\
\frac{\delta \mathcal{L}^{(0)}}{\delta e^{a}_{0}}&=&\Omega^{(0)}_{a},
\label{61a}
\end{eqnarray}

we obtain

\begin{eqnarray}
&&(-1)D_{0}\tilde{\theta}\frac{\partial \mathcal{L}^{(0)}}{\partial \omega_{0}} +
(-1)D_{0}\tilde{\theta}^{a}\frac{\partial \mathcal{L}^{(0)}}{\partial e^{a}_{0}} +
 (-1)\epsilon_{ab}\tilde{\theta}^{a}B^{b}\frac{\partial \mathcal{L}^{(0)}}{\partial B}+(\epsilon_{ab}\tilde{\theta}B^{b}-\epsilon_{ab}\tilde{\theta}^{b}B(1-2k^{2}B))\frac{\partial \mathcal{L}^{(0)}}{\partial B_{a}}+\nonumber \\ 
 &&(-1)(lD_{1}\tilde{\theta}^{a}-\epsilon{^{a}}_{b}\tilde{\theta}e^{b}_{1})\bigg(\frac{\partial \mathcal{L}^{(0)}}{\partial e^{a}_{1}}-\partial_{t}\frac{\partial \mathcal{L}^{(0)}}{\partial\dot{e}^{a}_{1}}\bigg)+ (-1)(\partial_{1}\tilde{\theta}+\frac{1}{l}\epsilon_{ab}\tilde{\theta}^{a}e^{b}_{1}(1-4k^{2}B))\bigg(\frac{\partial \mathcal{L}^{(0)}}{\partial \omega_{1}}-\partial_{t}\frac{\partial \mathcal{L}^{(0)}}{\partial\dot{\omega}_{1}}\bigg)=0.
\end{eqnarray}

Then, we can find that the gauge field transformations are given by 

\begin{eqnarray}
\delta{_{G}} B   &=& -\epsilon_{ab}\tilde{\theta}^{a}B^{b}, \nonumber\\
\delta{_{G}} B_{a}&=& \epsilon_{ab}\tilde{\theta}B^{b}-\epsilon_{ab}\tilde{\theta}^{b}B (1-2k^{2}B),\nonumber\\
\delta{_{G}} e^{a}_{0} &=& D_{0}\tau^{a}, \nonumber \\
\delta{_{G}} e^{a}_{1}&=& -lD_{1}\tilde{\theta}^{a} + \epsilon{^{a}}_{b}\tilde{\theta}e^{b}_{1},\nonumber \\
 \delta{_{G}} \omega_{0}&=& D_{0}\tau, \nonumber \\
\delta{_{G}}\omega_{1}&=& -\partial_{1}\tilde{\theta}-\frac{1}{l}\epsilon_{ab}\tilde{\theta}^{a}e^{b}_{1}(1-4k^{2}B).
\label{61}
\end{eqnarray}

It can be appreciated that the transformations obtained in equation $(\ref{61})$ coincide with those derived through the Dirac method in equation $(\ref{38})$.

\section{ Conclusions and prospects}

In this paper, we conduct a Hamiltonian canonical analysis on the action describing the Polynomial BF model system. Within the full phase space, we identify and distinguish all the constraints of the theory, classifying them into first and second classes.
Additionally, we determine the algebra of the first-class constraints, as shown in (\ref{28}) and count the degrees of freedom. The total degrees of freedom are calculated as $18$ (canonical variables) - $2 \times 6$ (first-class constraints) - $6$ (second-class constraints) = $0$, indicating a system with zero degrees of freedom. Moreover, we have removed the second-class constraints by introducing Dirac brackets, as we see from (\ref{33}). Castellani's approach enables us to deduce the fundamental gauge structure corresponding
to deformed $SO(3)$ transformations, as shown in  (\ref{38}). Specifying the gauge parameters allows us to derive diffeomorphisms and Poincaré symmetries originating from the fundamental gauge symmetry. On the contrary, treating the second-class constraints as strong equalities leads to outcomes consistent with the conventional expressions in the literature.

In comparison, we derive the structures of all the physical constraints in the Polynomial BF model using the modified FJ formalism. When we plug all obtained constraints into the Lagrangian through Lagrangian multipliers, we observe that the resulting two-form symplectic matrix remains singular, as shown in (\ref{51a}). This singularity implies that the investigated theory exhibits characteristics of a gauge system with additional symmetry. For effective quantization, we require suitable gauge conditions to modify the two-form symplectic matrix and render it non-singular. The modification enables us to derive basic quantum brackets from the inverse of the non-singular matrix. The transition amplitude can then be determined by assessing the determinant of the nonzero matrix. Additionally, we calculate the count of degrees of freedom in the theory, finding that it possesses zero degrees of freedom.
Moreover, by employing the MW algorithm in the FJ framework, we successfully obtain generators for gauge transformations and shifts throughout the entire configuration space, eliminating the need for Castellani’s algorithm.

Upon completing our calculations, we confirm the classical equivalence between Dirac brackets and FJ brackets, as obtained through the symplectic approach. In the FJ formulation, the process of identifying constraints is less intricate and involves fewer limitations compared to the Dirac method, rendering the FJ approach more elegant and efficient. 

Recently, it was demonstrated \cite{Pas} that the Polynomial BF model for gravity in two dimensions has employed a mechanism similar to that found in the MacDowell-Mansouri theory. Additionally, the Polynomial BF model, presented in \cite{Pas}, can be related to dilaton theories in two dimensions, as described in \cite{Val}. These theories play a fundamental role in the investigation of gravity in two dimensions. By incorporating a suitable dilaton field, it becomes possible to derive various gravitational models, including notable examples like the one proposed by Jackiw–Teitelboim and the quadratic gravitational formulation. The establishment of a connection between the Polynomial BF model and dilaton theory provides valuable insights for formulating an interaction term, facilitates the dynamic integration of torsion, and allows for an extension to four dimensions.



\section*{Declaration of competing interest}

The authors declare that they have no known competing financial interests or personal relationships that could have appeared to influence the work reported in this paper.

\section*{Acknowledgements}

J.M.C. welcomes the support of the Universidad Ju\'arez Aut\'onoma de Tabasco for providing a suitable work environment during the research period. J.M.C. also thanks CONACyT for their support through a grant for postdoctoral studies under grant No. 3873825.




\end{document}